\documentclass[12pt, twocolumn]{article}
\usepackage[top=1.2in,left=0.7in,footskip=0.5in,width=7.1in]{geometry}

\usepackage[utf8]{inputenc}

\usepackage{cite}

\usepackage{nameref,hyperref}
\hypersetup{
   colorlinks=true,             
   linkcolor=blue,              
   urlcolor=blue,               
   anchorcolor=blue,            
   citecolor=blue,              
}

\usepackage[right]{lineno}

\usepackage{microtype}
\DisableLigatures[f]{encoding = *, family = * }

\usepackage{changepage}

\usepackage[aboveskip=1pt,labelfont=bf,labelsep=period,singlelinecheck=off]{caption}

\usepackage{lastpage,fancyhdr}
\usepackage{epstopdf}

\usepackage{color}

\definecolor{Gray}{gray}{.25}

\usepackage{graphicx}

\usepackage{sidecap}

\usepackage{wrapfig}
\usepackage[pscoord]{eso-pic}
\usepackage[fulladjust]{marginnote}
\reversemarginpar

\usepackage{blindtext, multicol}
\usepackage{float}
\input{boxed1.sty}
\floatstyle{simplerule}
\restylefloat{figure}
\usepackage[table]{xcolor}
\definecolor{grey80}{rgb}{0.90,0.90,0.90}
\usepackage{amsmath}
\usepackage{natbib}
\begin{document}
   \onecolumn
   {\huge\textbf\newline{\bf Influence of Apophis' spin axis variations on a spacecraft during the 2029 close approach with Earth}}
   \newline
   \\
   S. Aljbaae\textsuperscript{1,*},
   J. Souchay\textsuperscript{2},
   V. Carruba\textsuperscript{3},
   D. M. Sanchez\textsuperscript{1},
   A. F. B. A. Prado\textsuperscript{1}\\
   \bigskip
   
   \noindent
   {\bf 1} Division of Graduate studies, INPE, C.P. 515, 12227-310 S\~ao Jos\'e dos Campos, SP, Brazil.\\
   {\bf 2} SYRTE, Observatoire de Paris, PSL Research University, CNRS, Sorbonne Universités, UPMC Univ. Paris 06, LNE, 61 avenue de l'Observatoire, 75014 Paris, France.\\
   {\bf 3} S\~{a}o Paulo State University (UNESP), School of Natural Sciences and Engineering, Guaratinguet\'{a}, SP, 12516-410, Brazil.
   \\
   \bigskip
   
   \noindent
   * \href{safwan.aljbaae@gmail.com}{safwan.aljbaae@gmail.com}

   \section*{Abstract}
      Tumbling asteroids belong to a small group of objects, whose angular velocity vector is unaligned with any of its principal axes of inertia. This leads to challenging efforts to model the trajectory of any spacecraft designed to orbit these bodies. In this work, we deepen a previous study on this topic, concerning the asteroid (99942) Apophis during its close encounter with the Earth in 2029. We analyze the orbital behaviour of a spacecraft orbiting the asteroid during this event, by including the effects of the changes of orientation of the spin axis of the asteroid, depending on two sets of initial conditions. The global dynamics of the spacecraft around the target are analyzed using three approaches, MEGNO, PMap, and Time-Series prediction. We confirm that no spacecraft with natural orbits could survive the high perturbations caused by the close encounter with our planet.\\
      
      {\bf Key words:} Celestial mechanics — minor planets, asteroids: individual (Apophis) — gravitation. \\
   \twocolumn      
   \section{Introduction}\label{sec01_introduction}
\begin{figure}[!ht]
         \includegraphics[width=\linewidth]{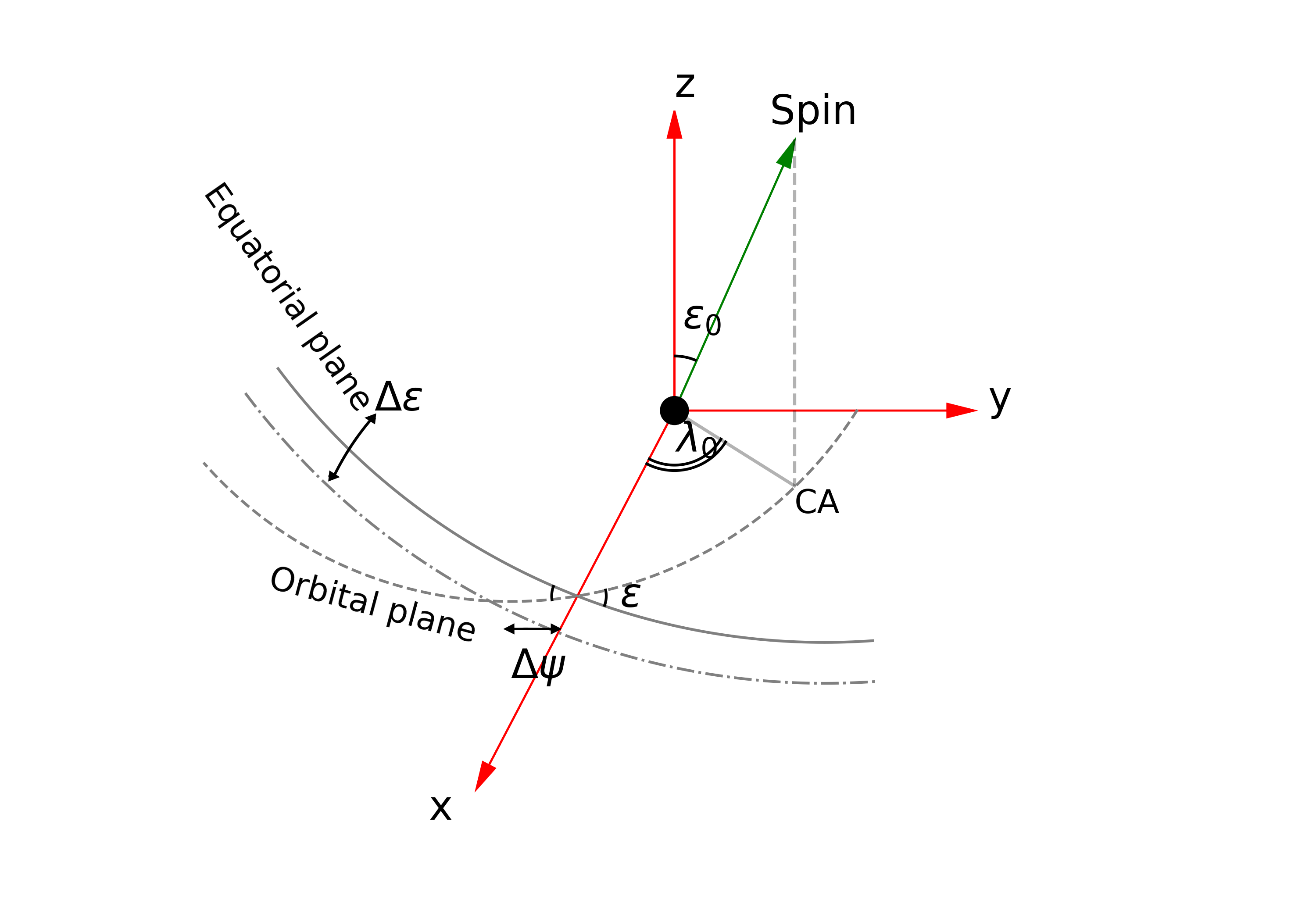}
         \caption{Parametrization of the spin axis of Apophis during the Earth close encounter.} \label{fig01_obliquity_precession}
      \end{figure}

      The near-Earth asteroid 99942 Apophis is a small 387 meters body with a surface gravity of about 0.0023 cm/s$^2$ \citep{pravec_2014, muller_2014}. This asteroid is classified as potentially dangerous in the future and it will experience a close encounter with Earth on 2029 April 13, at about 38 000 km from the Earth's center. A mission that places a spacecraft in orbit around (99942) Apophis just before the close encounter with our planet would be very important because, for the first time, it could make close-up observations of an asteroid during a close encounter with our planet, which can change the orientation of the spin axis of the target, as well as its shape. In fact, the encounter of Apophis with our planet excites dynamically the motion of any particle located in a zone around the asteroid, possibly causing its collision or escape from the target, as shown by \citet{aljbaae_2021}. In this last study, possible changes of orientation of the spin axis of the target were neglected. On the contrary, in this paper, our aim is to provide a more complete dynamical analysis of an orbiting spacecraft including these changes. Notice that Apophis is well known to be in a moderatly tumbling rotation state \citep{pravec_2014}, and that the Earth gravitational potential will affect significantly the spin state of the target. \citet{souchay_2018} estimated the changes of the angular momentum axis of Apophis, whose the direction is considered to be very close to the axis of rotation. These changes were calculated starting from the two fundamental parameters representing the motion of Apophis axis in space. These are $\varepsilon$, the obliquity angle between the orbital and equatorial plane of the asteroid, and $\psi$, the precession angle, which expresses the motion of the ascending node of the equator with respect to the orbital plane (Fig. \ref{fig01_obliquity_precession}). 
      
      In this paper, we aim at characterizing the effects of the changes of the spin axis and rate of (99942) Apophis on the motion of a spacecraft orbiting the asteroid. Our work is divided as follows. In Sec. \ref{sec02_spin_axis_variations} we use the equations of motion of \citet{souchay_2018}, to recalculate the extreme amplitudes of the effects of the close approach on the Apophis spin axis, according to the range of adequate initial conditions. Then, a complete dynamical model of a spacecraft in orbit around the target is presented in Sec. \ref{sec03_Dynamical_model}, where we take into account these last effects. In Sec. \ref{sec04_spin_effects}, we calculate the specific effects of the changes of the spin axis of Apophis on the orbit of the spacecraft, depending on two sets of initial conditions. In Sec. \ref{sec05_stability_region}, we consider the full set of perturbations to characterize the dynamical evolution of a spacecraft around Apophis. We investigate a new approach based on Time-Series Prediction with Machine Learning to identify non-chaotic regions. We compare the results of this new approach to those we obtain from PMap \citep{sanchez_2017, sanchez_2019} and MEGNO \citep{cincotta_2000} methods using nearby orbits.
   
   \section{Variations of Apophis spin axis}\label{sec02_spin_axis_variations}
   
      \citet{pravec_2014} used a large set of lightcurves observations to determine the rotational state of Apophis. Their conclusion was that the brightness of Apophis did not repeat with a single period, but it showed the characteristics of a slightly tumbling rotational state. Using a 2-period Fourier series method, they found two components, the main one, with period $P_1 =$ 30.56h, which corresponds to a leading rotation component around the SAM(Short Axis Mode) and a secondary one with a period around $P_2=$ 29.05h, but with a relatively big uncertainty. Nevertheless, the amplitude of the wobble around the SAM is rather small, between $20^{\circ}$ and $25^{\circ}$, which characterizes a moderate excitation. Whatever be the amplitude of the wobble during the close encounter, we do not take into account the small changes of the orientation of the spin axis which should occur in the case of the absence of the close encounter. We concentrate on the relatively large changes due to the presence of the Earth during the close encounter. They are presented in the following.

\begin{figure}[!ht]
         \includegraphics[width=0.49\linewidth]{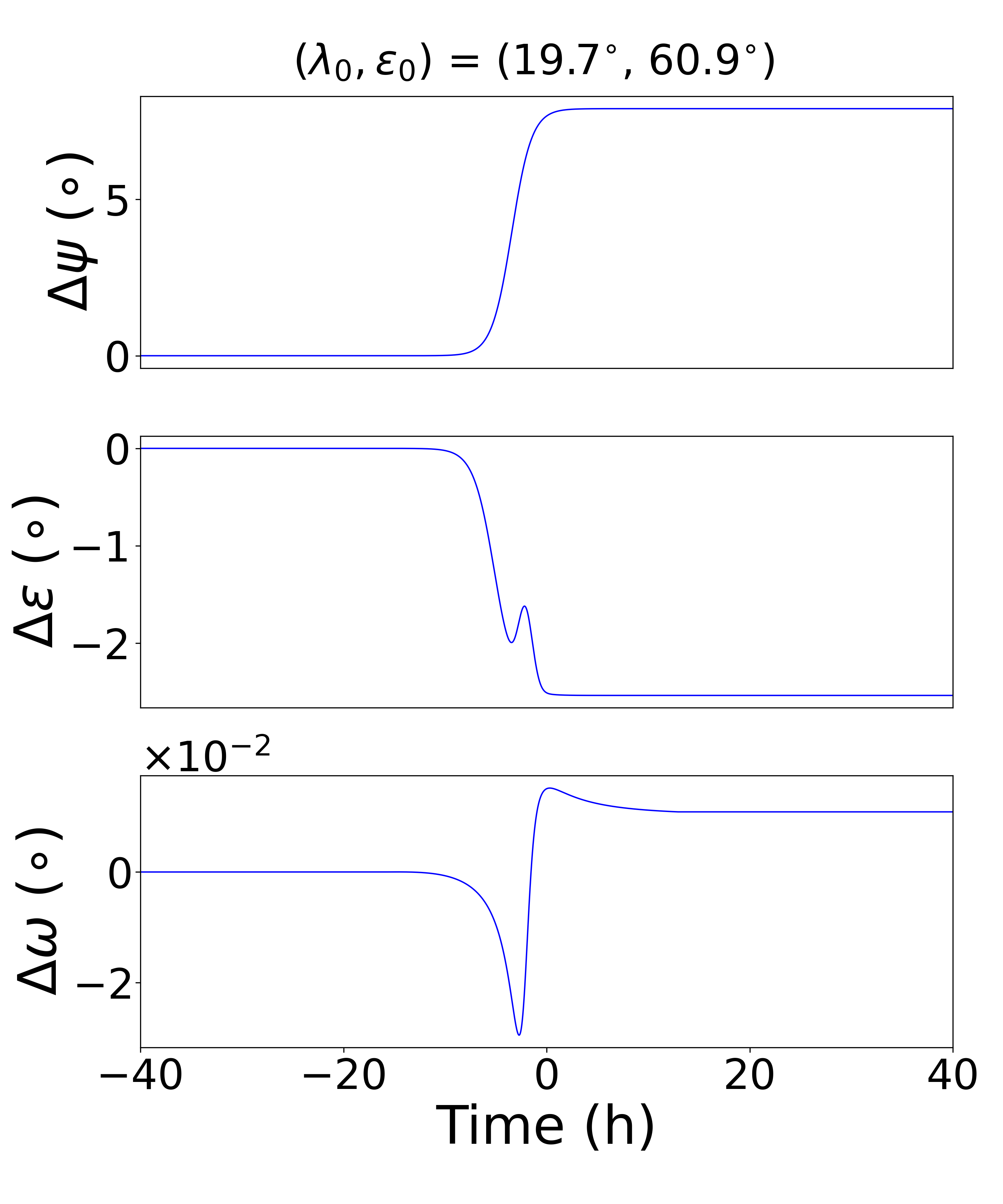}
         \includegraphics[width=0.49\linewidth]{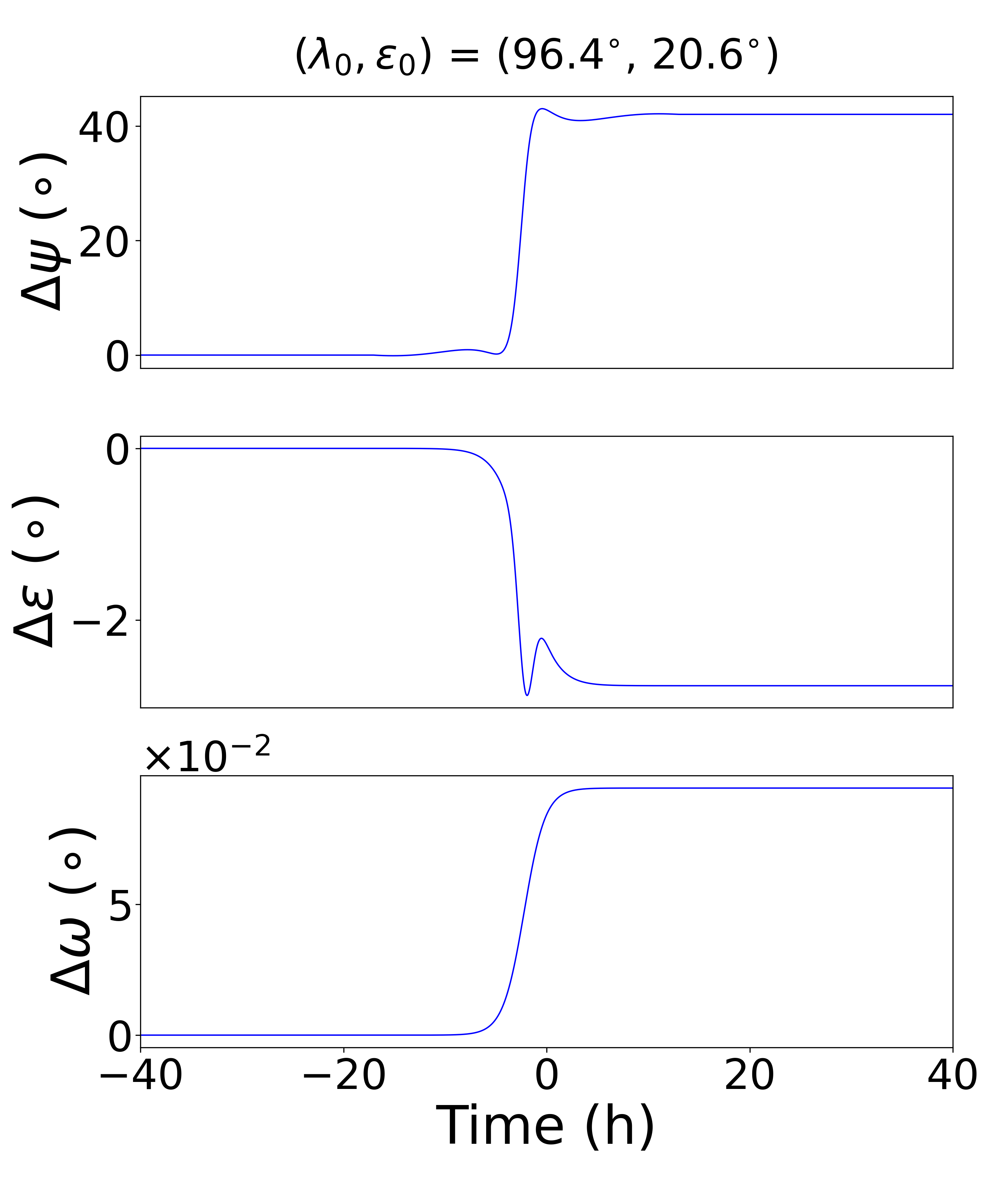}
         \caption{Minimum (left column) and Maximum (right column) variations ofthe orientation of Apophis spin axis.} \label{fig02_plot_delta_rot}
      \end{figure}

      Changes in the spin-state of Apophis were estimated by \citet{scheeres_2005} conducting Monte-Carlo simulations modeling the asteroid as a tri-axial ellipsoid with a length-to-width ratio of 1.4. More recently, \citet{souchay_2018} used the classical theory of the rotation of the Earth, originally developed in \citet{kinoshita_1977}, to calculate the variations of the orientation of Apophis angular momentum axis during the close encounter due to the tidal deformation associated with the gravitational potential of our planet. The authors used the physical characteristics of the asteroid from \citet{pravec_2014}. Thus they showed that the values of the spin axis orientation parameters just before the encounter should be bounded in the intervals [$10^{\circ}$ - $100^{\circ}$] for $\lambda_{0}$ and [$10^{\circ}$ - $70^{\circ}$] for $\varepsilon_{0}$, according to the observational data from \citet{pravec_2014}. They tested numerous initial values of ($\lambda_{0}$, $\varepsilon_{0}$) and showed that the asteroid could undergo dramatic changes in obliquity ($\varepsilon$) and a longitude of the ascending node ($\psi$). The formulas leading to the determinations of these changes are given in Appendix \ref{appendix_1}. In this work, we choose to apply the study of \citet{souchay_2018} to determine the minimum and maximum values of the amplitudes of changes of the orientation of the spin axis, to investigate the effects of these changes on the dynamics of a spacecraft in orbit about Apophis during the close encounter. The minimum and maximum values of the variations are obtained for the pairs ($\lambda_{0}$, $\varepsilon_{0}$) = (19.7$^{\circ}$, 60.9$^{\circ}$) and (96.4$^{\circ}$, 20.6$^{\circ}$), respectively, as they are shown in Fig. \ref{fig02_plot_delta_rot}. Our results are slightly different from the ones found by \citet{souchay_2018}, because we use a slightly different and more recent set of physical parameters of the asteroid, determined by \citet{brozovic_2018}.      
      
   \section{Dynamical model}\label{sec03_Dynamical_model}
   
      In the following, we investigate the motion of a spacecraft in orbit around Apophis by taking into account the variations of the spin axis of the target during the close encounter, as calculated and described above. The equations of motion used here are referred to an inertial reference frame with origin on the centre of mass of the asteroid, with two axes along the J2000 Ecliptic plane. As done in \citet{aljbaae_2021}, we consider the asteroid as a cloud of 3996 point masses, corresponding to the number of faces in the shape derived from \citet{brozovic_2018} to calculate the gradient of its gravitational potential. Concerned readers can find a complete discussion about the validation of this gravitational potential modelling in \citet{aljbaae_2021}. To reach the accuracy of the orientation and distance from the Earth to Apophis provided by the JPL's HORIZONS ephemerides\footnote{\href{https://ssd.jpl.nasa.gov/?horizons}{https://ssd.jpl.nasa.gov/?horizons}}, we take into account the gravitational influence of the Sun, the eight planets, the Moon, Pluto, and the three largest asteroids, Ceres, Pallas, and Vesta. The initial conditions for all the bodies are provided by the JPL's HORIZONS ephemerides on March 1, 2029. Our determination for the minimum distance during the encounter is 37723 km, which fits very well with the value of 37728 km given by HORIZONS. The gravitational potential of the Earth and of the Moon are expanded using the spherical harmonics up to degree and order 4, as implemented in \citet{sanchez_2017} and \citet{sanchez_2014}. Our dynamical model also includes the Solar Radiation Pressure (SRP), as described in \citet{beutler_2005}. We apply this effect for an OSIRIS-REx-like spacecraft with a reflectance of 0.4 and a mass-to-area ratio of 60 kg.m$^{-2}$. Thus the equations of motion for a spacecraft close to (99942) Apophis are given by:
      
      \begin{eqnarray}\label{motion1}
         \ddot{\text{r}}&=& U_{\text{r}} + \sum_{i=1}^{14}\mathcal{G}m_{i}\big(\frac{\text{r}_{i}-\text{r}}{|\text{r}_{i}-\text{r}|^{3}} - \frac{\text{r}_{i}}{|\text{r}_{i}|^{3}}\big) + \nonumber \\
         &&\text{P}_{\text{E}} + \text{P}_{\text{M}} + \nu\text{P}_{\text{R}}
      \end{eqnarray}

      \noindent where, $\text{r}$ is the position vector of the spacecraft in the inertial frame of reference, $\text{r}_{i}$ and $\mathcal{G}m_{i}$ are the position vector and gravitational parameter of the $i^{\text{th}}$ body, with $\mathcal{G} = 6.672 59 \times 10^{-20}$ km$^{3}$kg$^{-1}$s$^{-2}$. $\text{P}_{\text{E}}$ and $\text{P}_{\text{M}}$ are, respectively, the acceleration due to the gravitational potential of the Earth and of the Moon, described by the spherical harmonics up to degree and order 4. $\nu\text{P}_{\text{R}}$ represents the acceleration due to the direct radiation pressure considering the shadowing phenomenon, as described in our previous work \citep{aljbaae_2021}. $U_{\text{r}}$ is the gradient of the gravitational potential of asteroid, calculated from a sum of 3996 points after rotating the shape \citet{pravec_2014} about the origin, in terms of longitude, obliquity, the rotation period $P_{\omega} = 30.4$ h and the precession period $P_{\psi} = 263$ h. In order to orientate Apophis with respect to our reference frame, we apply a sequence of rotations that can be represented as follows:
      \begin{eqnarray}\label{rotations}
          && R_{z}(\frac{2\pi}{p_{\psi}}t+\lambda_{0}+\Delta\psi)\nonumber\\
          && R_{x}(\varepsilon_{0}+\Delta\varepsilon) \\
          && R_{z}(\frac{2\pi}{p_{\omega}}t + \Delta\omega)\nonumber
      \end{eqnarray}
      \noindent where, $R_{x}$ and $R_{z}$ are the rotation matrix about the x-axis and z-axis, respectively. $\Delta\psi, \Delta\varepsilon$, and $\Delta\omega$ are the variations of the Apophis axis as calculated in the previous section. Here, we suppose that the shape of the asteroid and the periods $p_{\omega}$, $p_{\psi}$ does not change significantly during the Apophis/Earth close encounter. The estimation of these changes is still an open question that needs to be investigated. \\
      

   \section{Effects of the spin axis variations on an orbit around Apophis}\label{sec04_spin_effects}

\begin{figure}[!htp]
         \centering{$a_{0} = 0.45$ km}\\
         \includegraphics[width=0.47\linewidth]{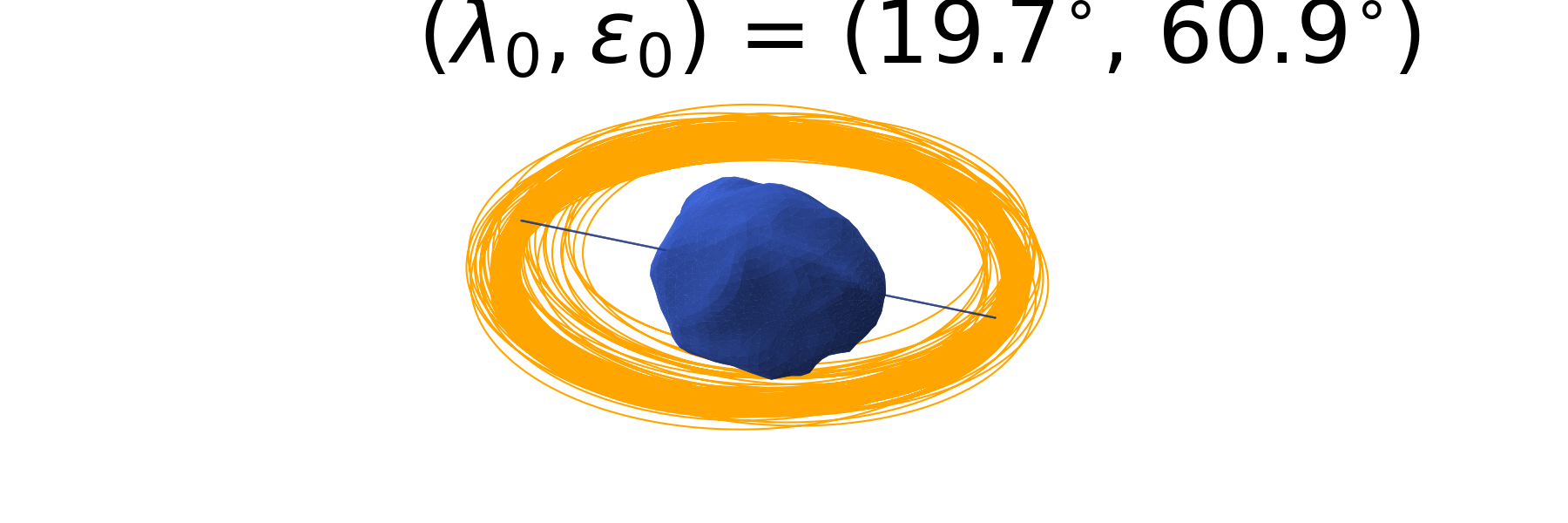}
         \includegraphics[width=0.47\linewidth]{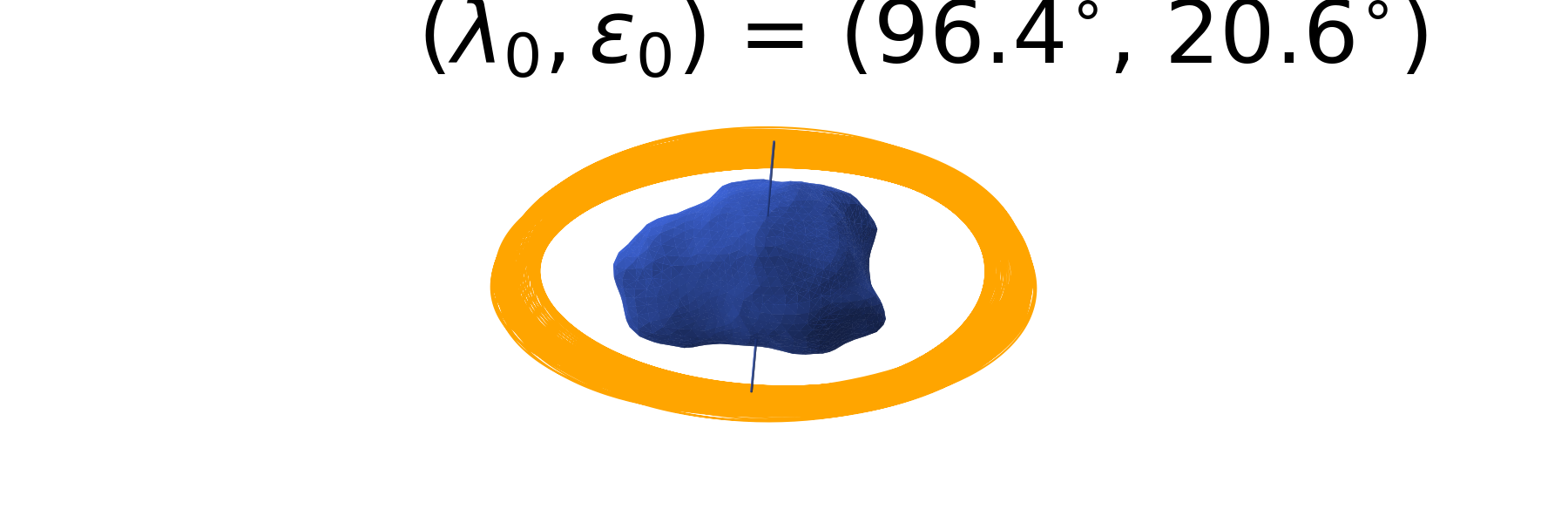}\\
         \includegraphics[width=0.47\linewidth]{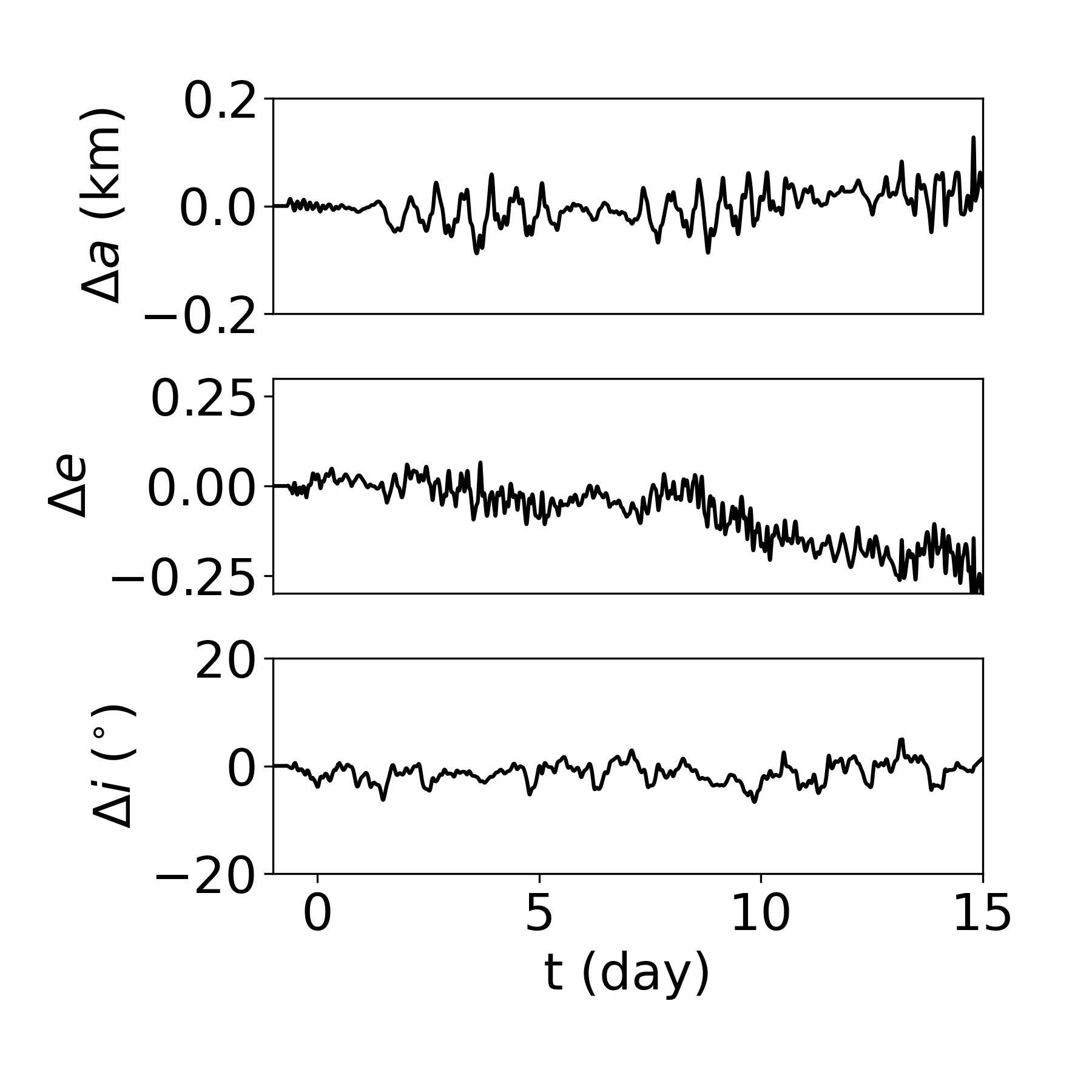}
         \includegraphics[width=0.47\linewidth]{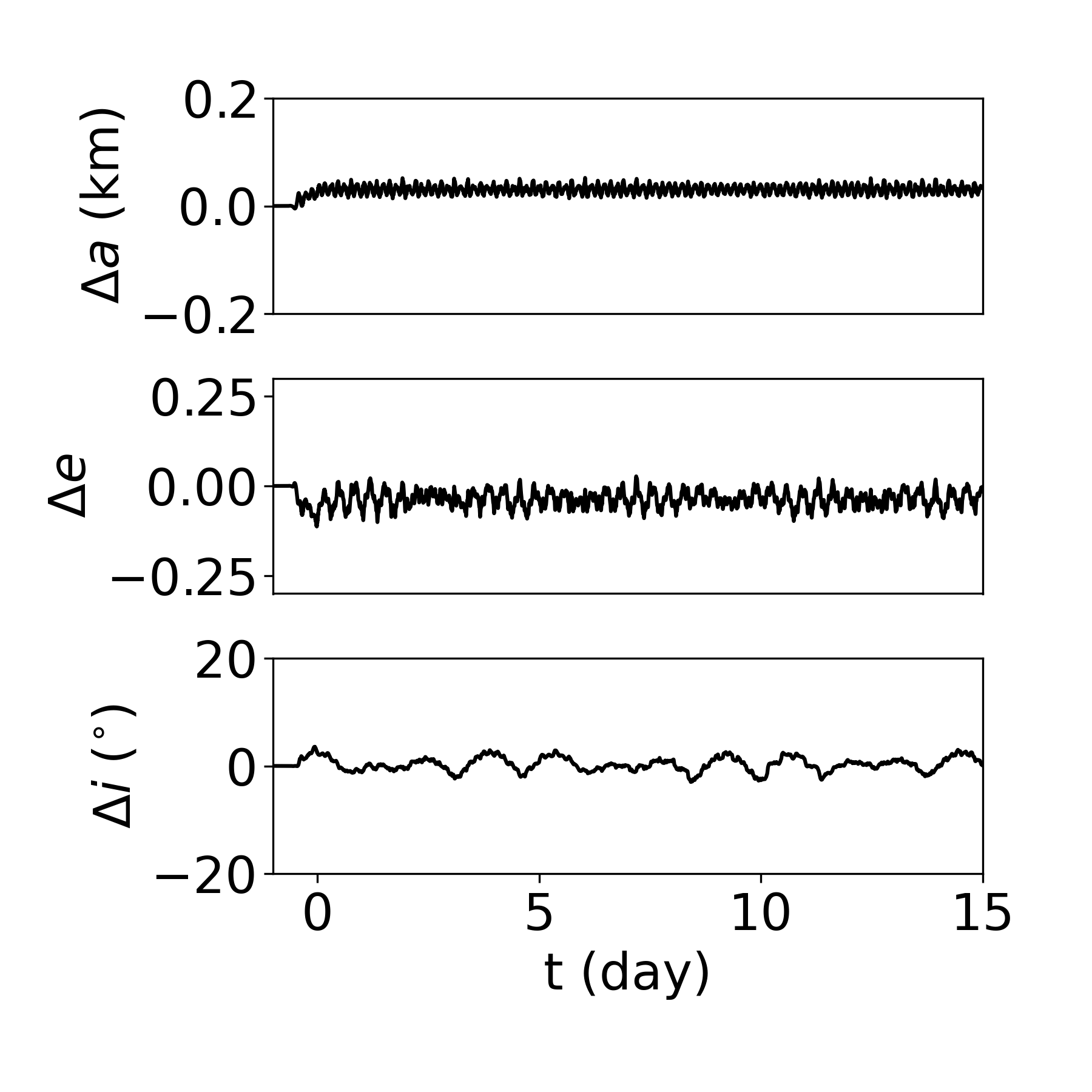}\\
         \centering{$a_{0} = 0.50$ km}\\ 
         \includegraphics[width=0.47\linewidth]{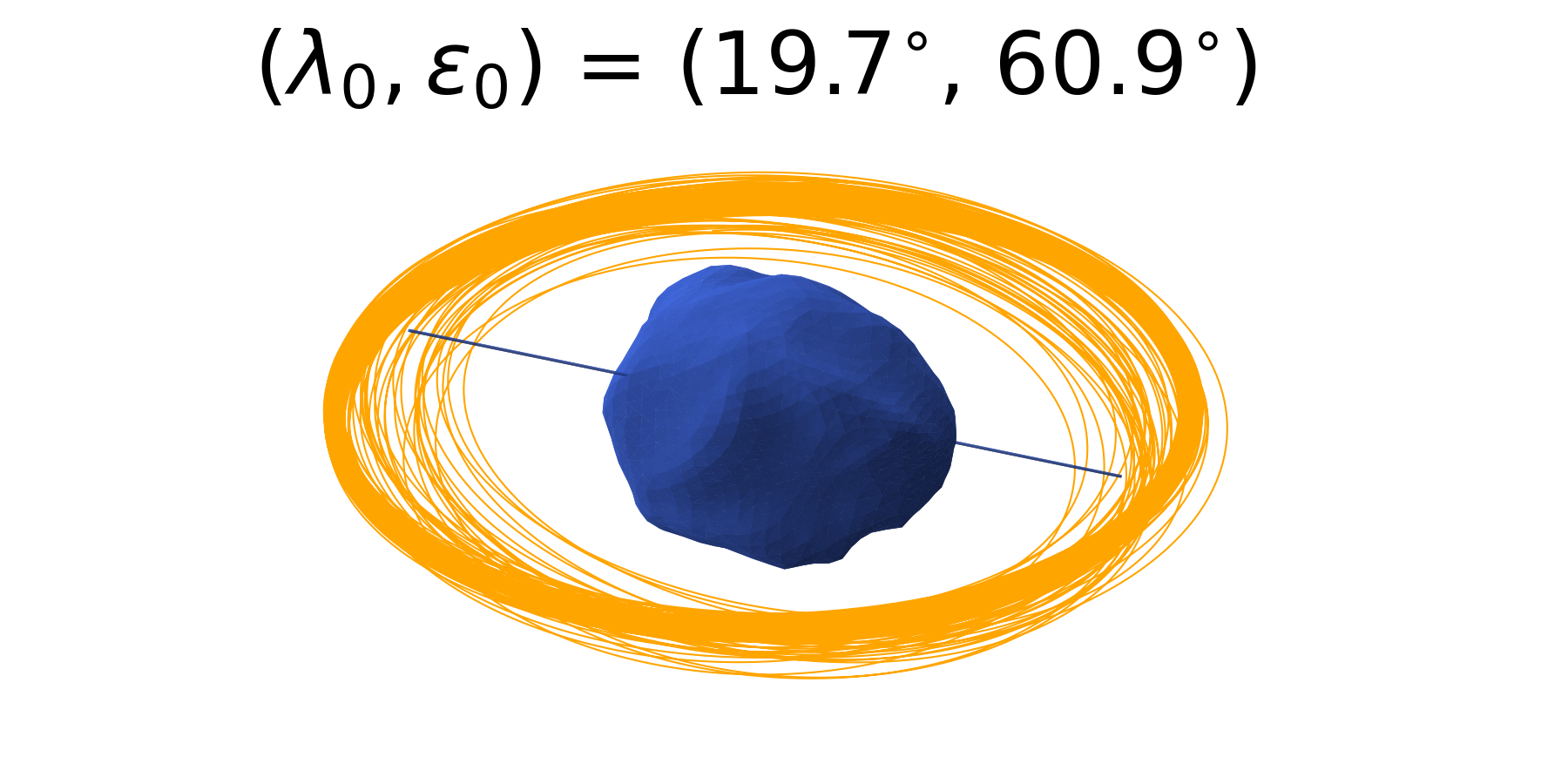}         
         \includegraphics[width=0.47\linewidth]{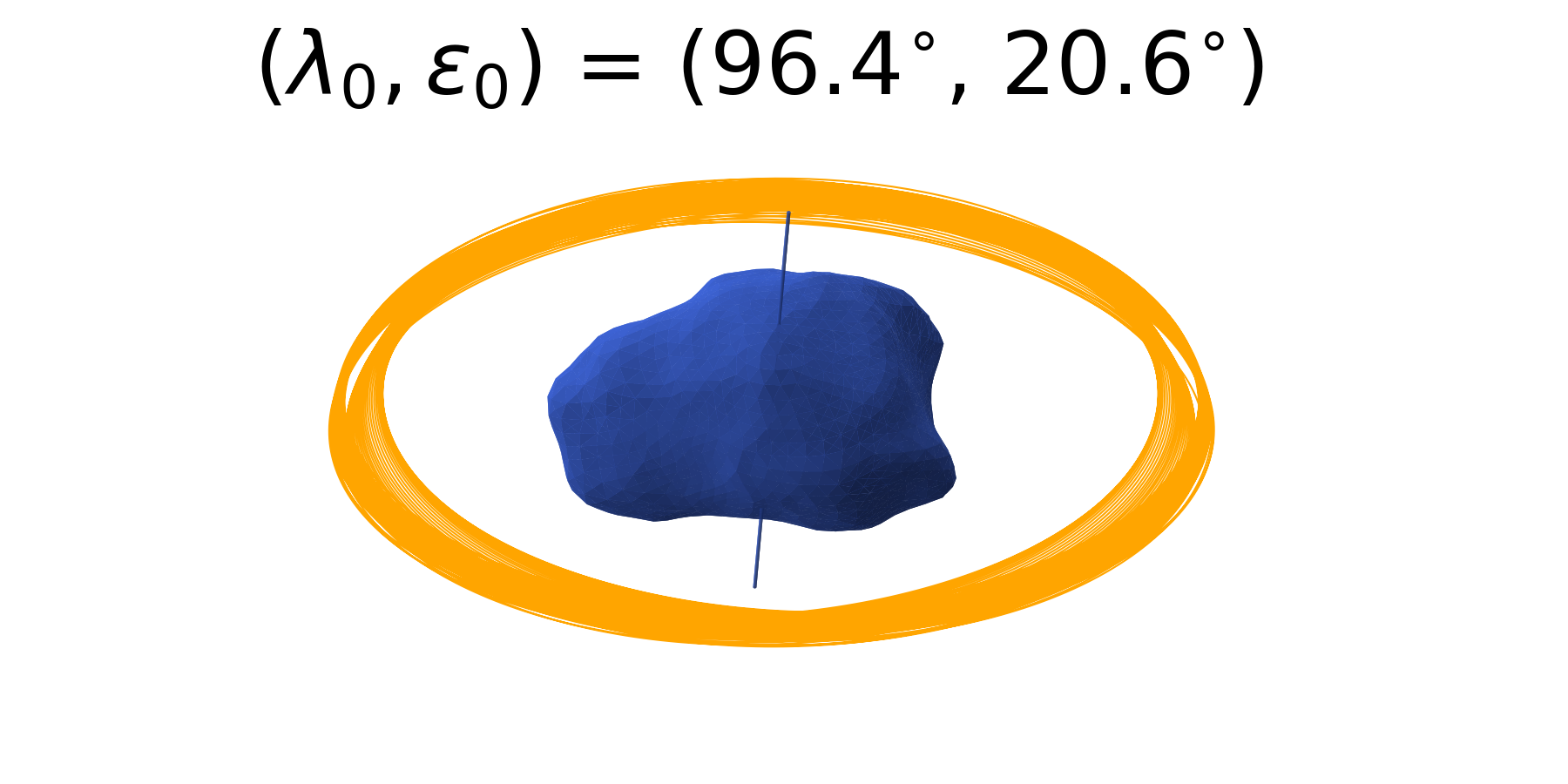}\\
         \includegraphics[width=0.47\linewidth]{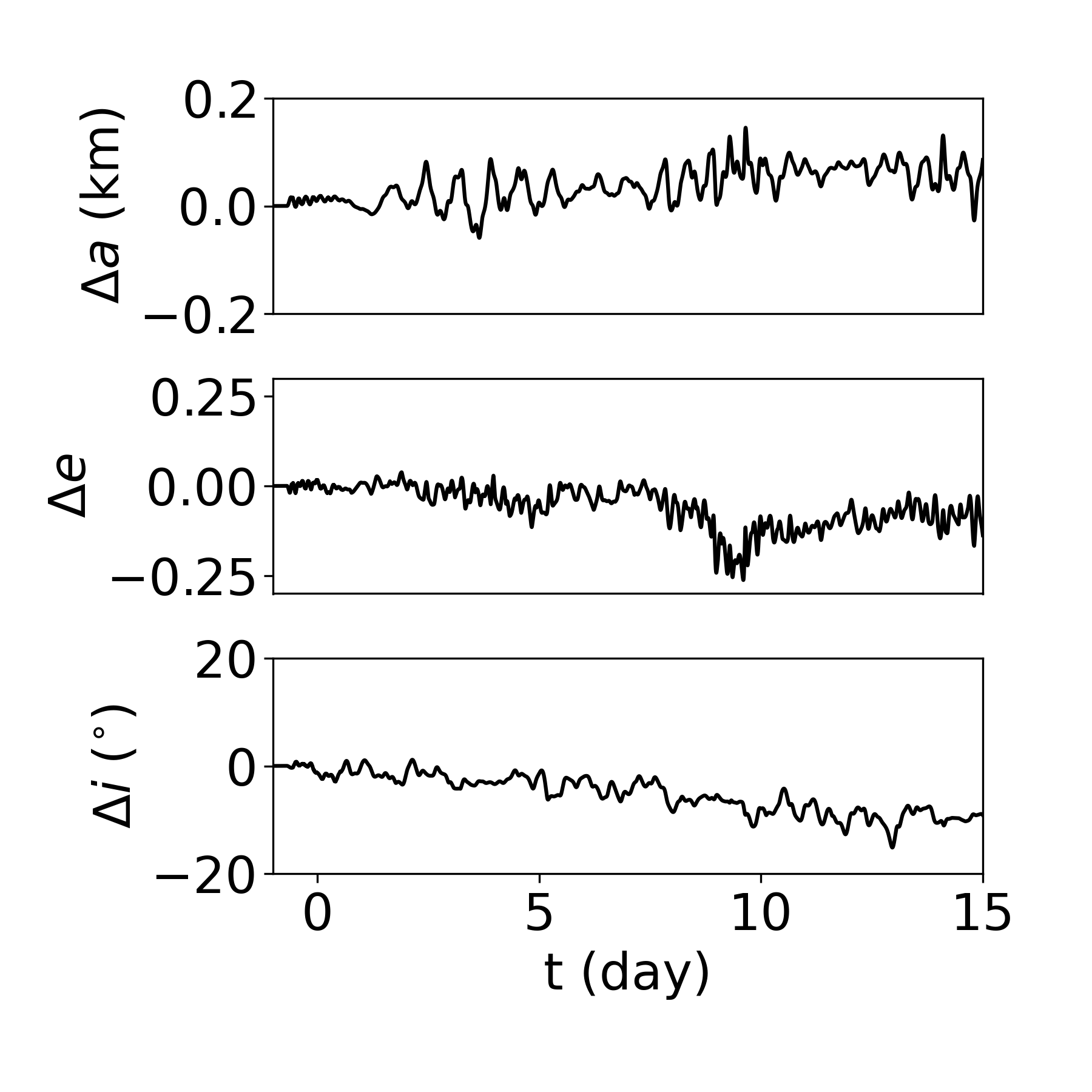}
         \includegraphics[width=0.47\linewidth]{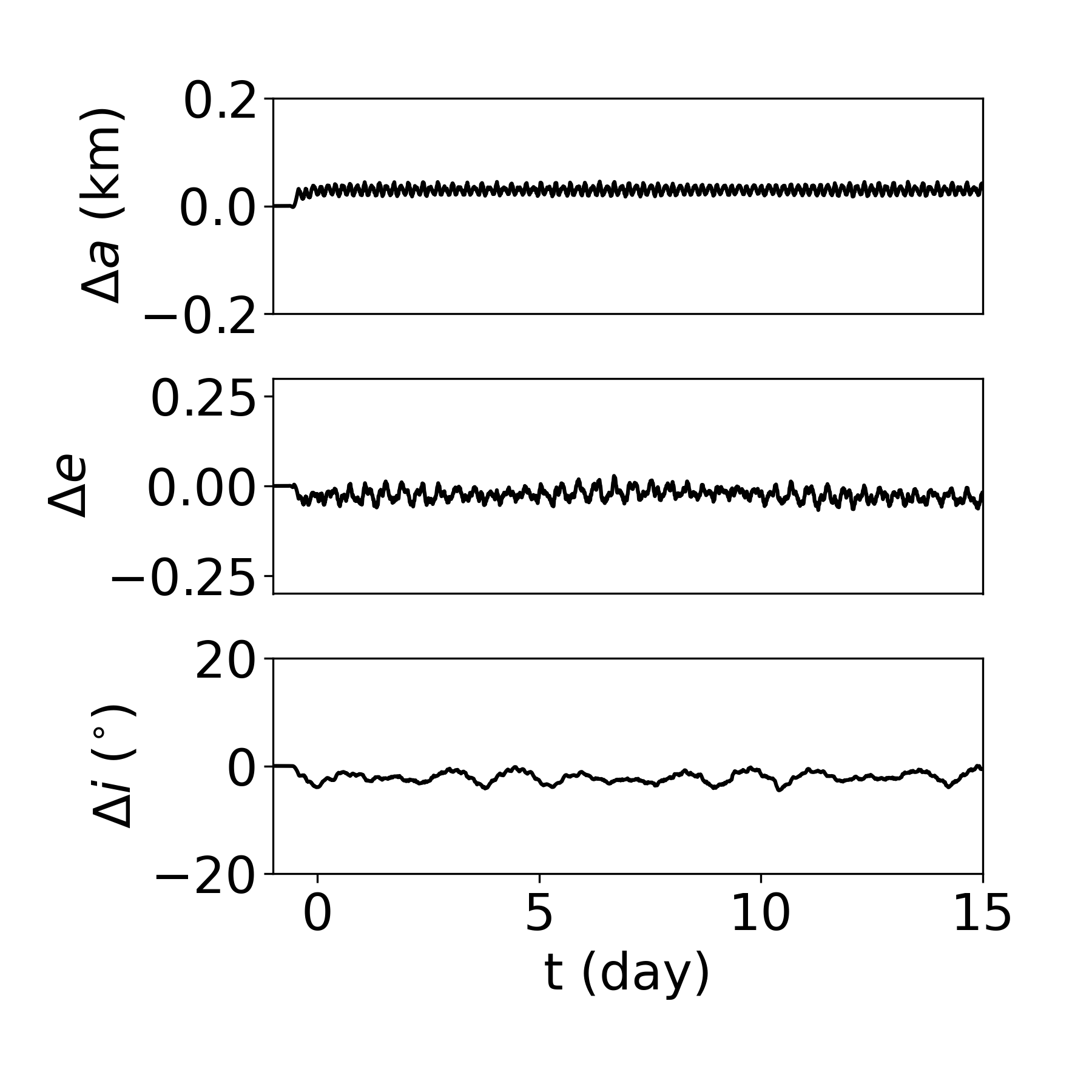}\\
         \centering{$a_{0} = 1.00$ km}\\ 
         \includegraphics[width=0.47\linewidth]{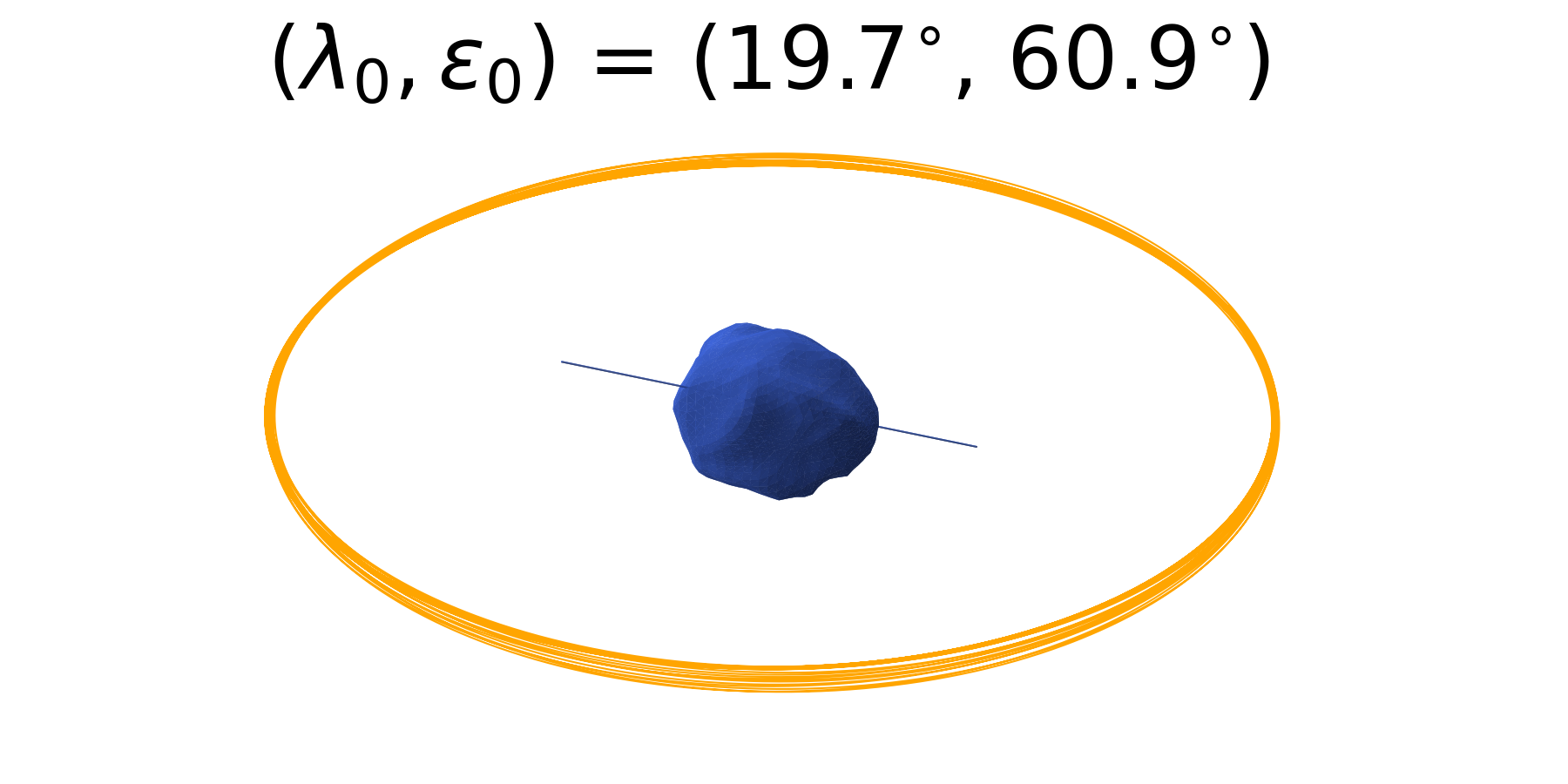}         
         \includegraphics[width=0.47\linewidth]{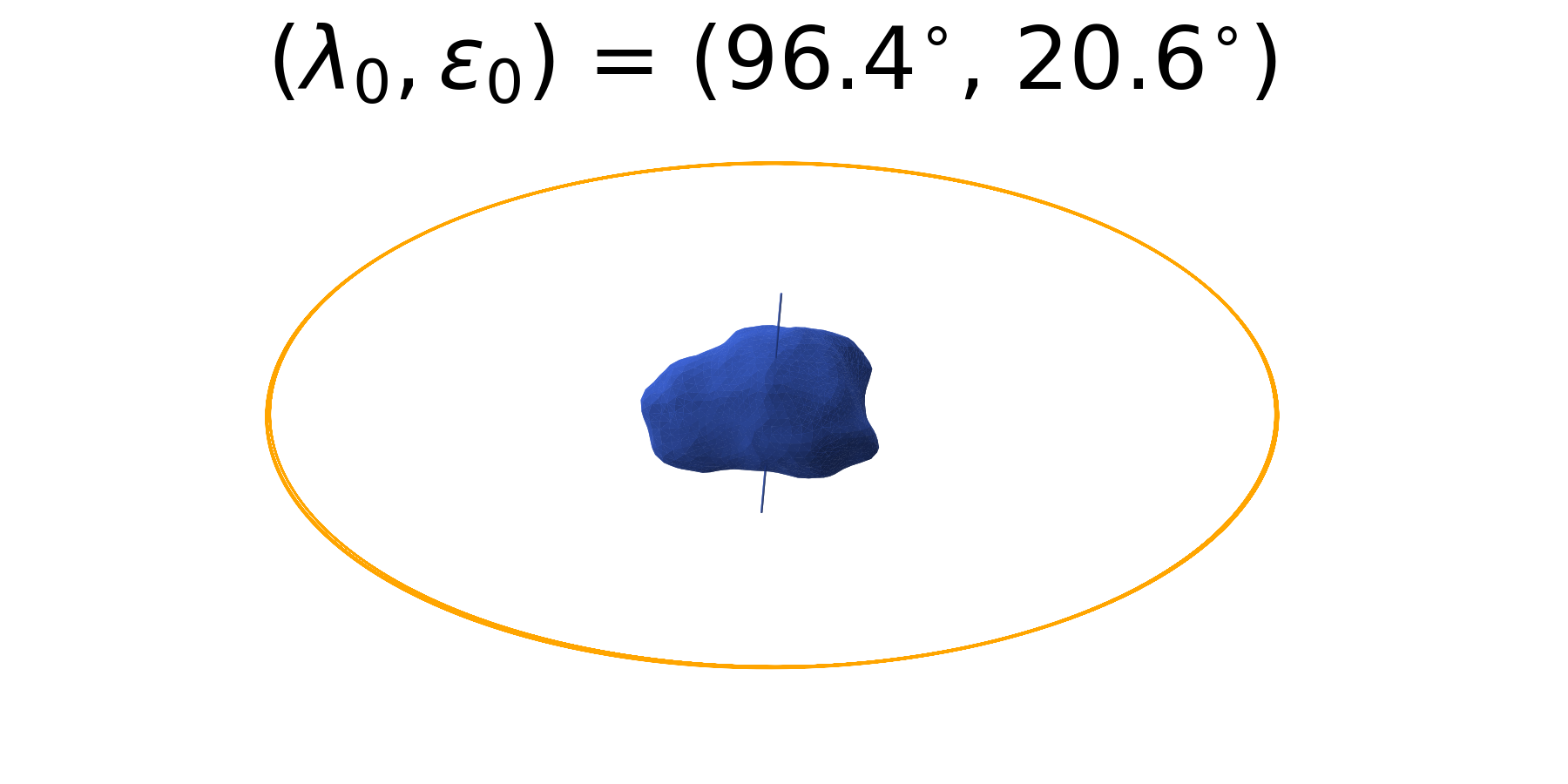}\\
         \includegraphics[width=0.47\linewidth]{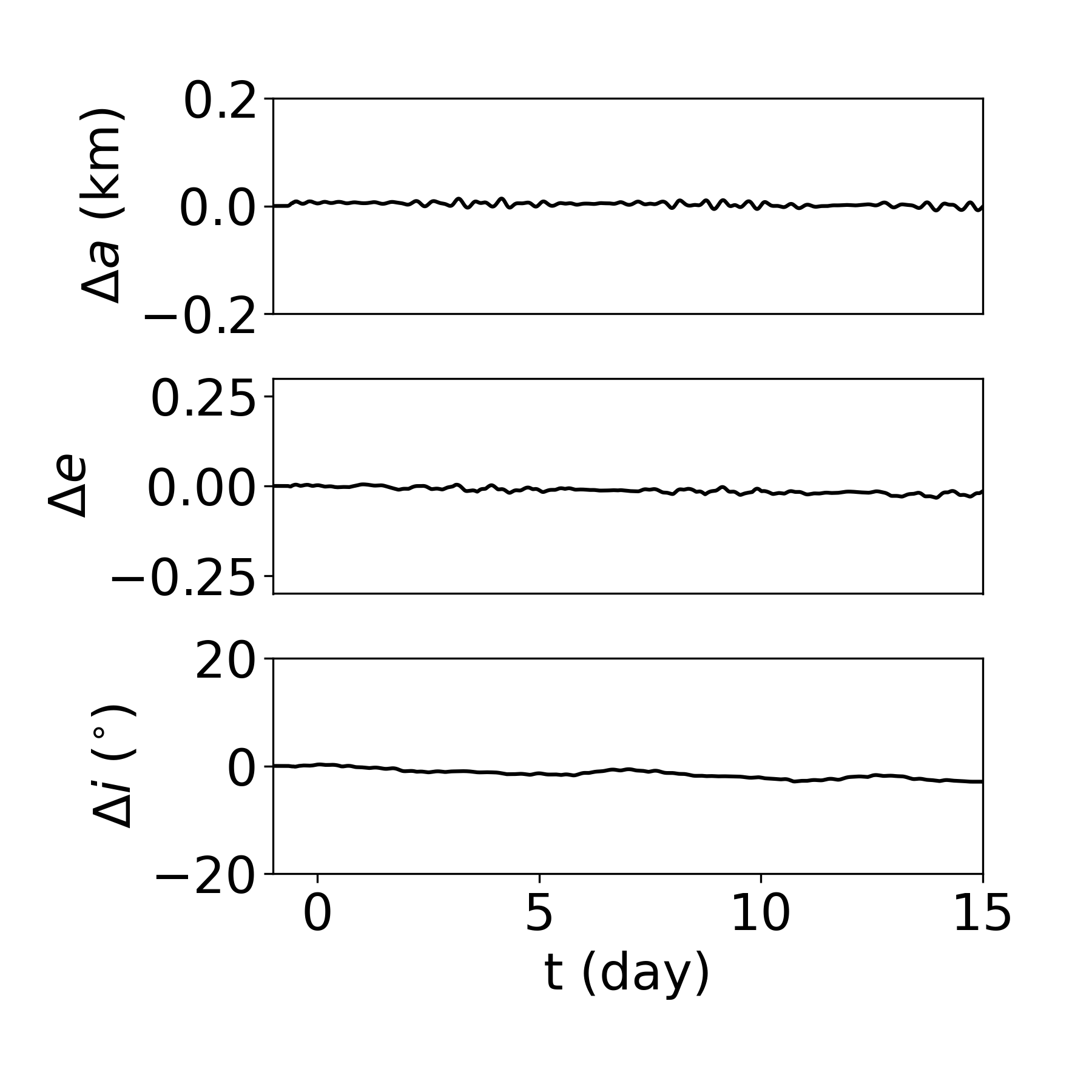}
         \includegraphics[width=0.47\linewidth]{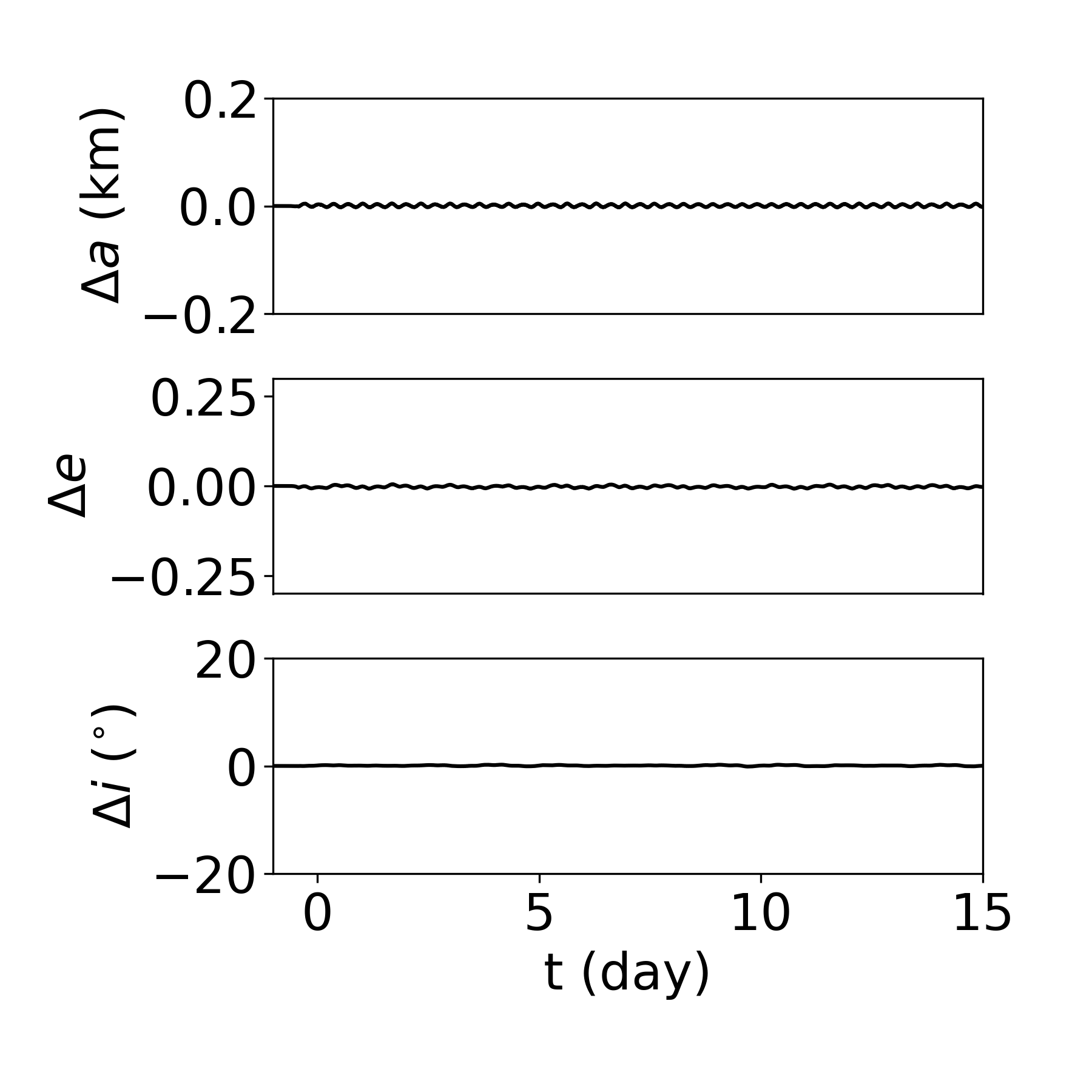}        
         \caption{Effects on $a$, $e$ and $i$ of the changes of Apophis spin axis during the Earth close encounter, on orbits with $a=0.45, ~ 0.5, ~ 1.0$ km. All the other orbital parameters are fixed to 0.} \label{fig03_spin_effects_orbit}
      \end{figure}

\begin{figure}[!ht]
          \includegraphics[width=1\linewidth]{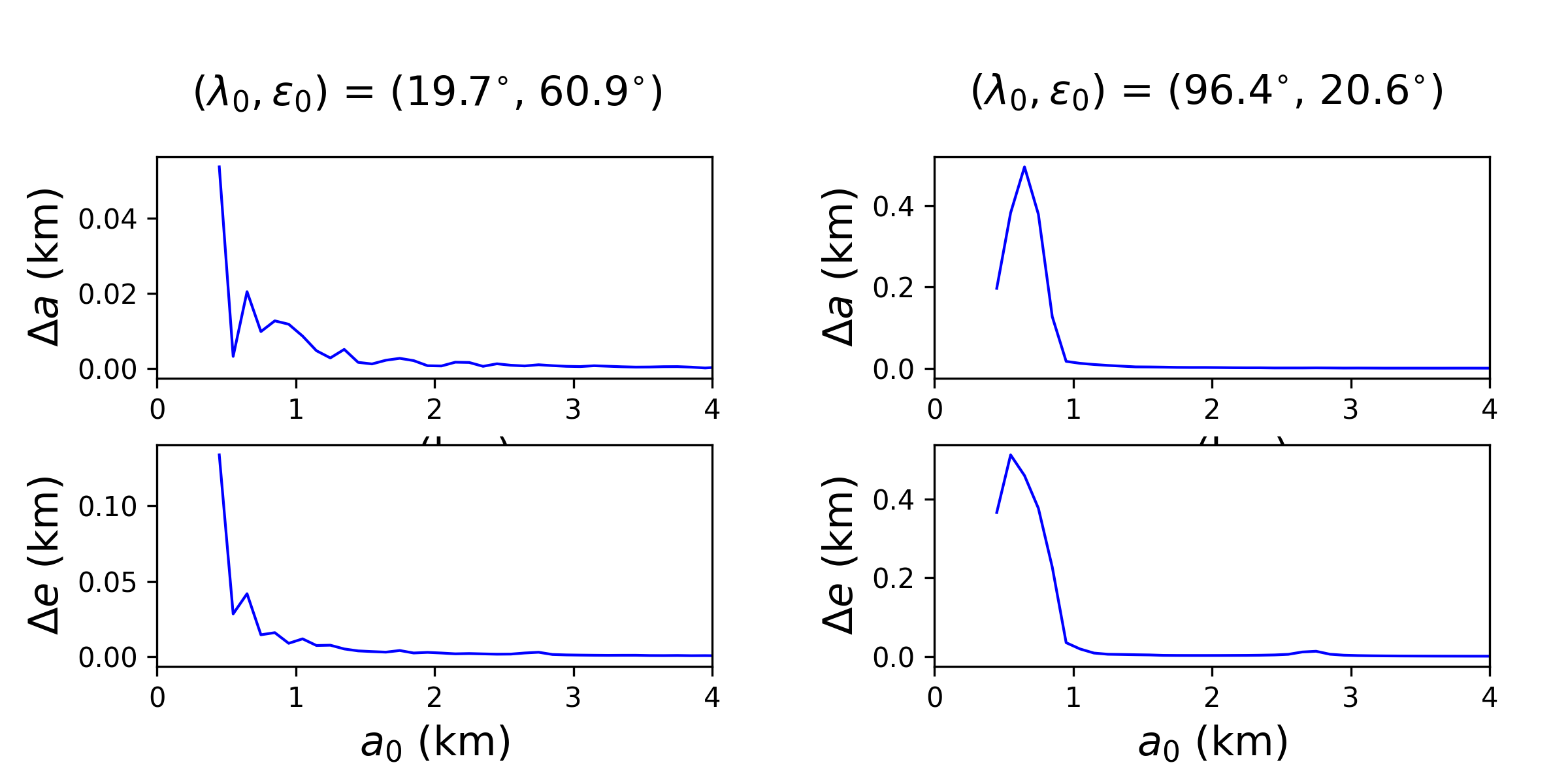}
          \caption{Peak-to-peak amplitudes of the variations of semi-major axis ($\Delta a$) and eccentricity ($\Delta e$) of a spacecraft due to the variations of the spin axis of Apophis, with respect to the initial value of the semi-major axis.} \label{fig04_axis_effects_energy}
      \end{figure}

      In this section, we investigate the specific effects of the variations of orientation of Apophis spin axis on the orbit of the particle (spacecraft). We consider here the initial values for the minimum and the maximum of these variations, as outlined in section \ref{sec02_spin_axis_variations}. As we are concerned here by the very short time interval of the close encounter, that is to say a few hours, we do not take into account the gravitational effect of the other bodies of the Solar system and of the SRP. They are omitted just to isolate the effects of the changes of the Apophis spin axis. We consider an initial circular retrograde orbit along the equatorial plane of the asteroid ($i=0^{\circ}$), by testing successive values of the semi-major axis ($a_{0}$) of 0.45, 0.5 and 1.0 km. Orbits with $a_{0} < $ 0.45 km will collide with the central body and should be investigated with a more precise dynamical model such as the classical polyhedral approach \citep{werner_1997, tsoulis_2001}, which is out of the scope of this work. Our results are presented in Fig. \ref{fig03_spin_effects_orbit}. We can notice that the minimum spin variations ($\lambda_{0}$ = 19.7$^{\circ}$, $\varepsilon_{0}$ = 60.9$^{\circ}$) generate bigger effects on the orbits than the maximum spin variations ($\lambda_{0}$ = 96.4$^{\circ}$, $\varepsilon_{0}$ = 20.6$^{\circ}$), as we will see later of on this work. In Table \ref{tabe_01_effects_spin} we list the peak-to-peak amplitude of the variations of the orbital elements in each case, as well as the minimum and maximum period of significant components calculated by the fast Fourier transform (FFT). We remark that these effects are considerably attenuated as the orbit is away from the asteroid, as shown in Fig. \ref{fig04_axis_effects_energy}, where we plot the peak-to-peak amplitudes of the variations of semi-major axis ($\Delta a$) and of the eccentricity ($\Delta e$) of the spacecraft due to the variations of the spin axis, as a function of the initial value of the semi-major axis. We remark that the values are quite different according to the initial conditions for the Apophis spin axis orientation.
      
      \begin{table}
      \begin{minipage}[!htp]{1.0\linewidth}
      \caption{Effects on $a$, $e$ and $i$ of the changes of Apophis spin axis during the Earth close encounter, on orbits with $a=0.45, ~ 0.5, ~ 1.0$ km. All the other orbital parameters are fixed to 0.}\label{tabe_01_effects_spin}
      \begin{center}
      \resizebox{0.99\textwidth}{!}{
      \begin{tabular}{|l|rrr|rrr|}
         \hline
         & \multicolumn{3}{c}{$\lambda_{0}$ = 19.7$^{\circ}$, $\varepsilon_{0}$ = 60.9$^{\circ}$} & \multicolumn{3}{|c|}{$\lambda_{0}$ = 96.4$^{\circ}$, $\varepsilon_{0}$ = 20.6$^{\circ}$} \\
         & amp. & Min. Friq. &   Max. Friq   &  amp. & Min. Friq. &   Max. Friq\\
         &      & (days)     & (days)        &       & (days)     & (days)\\
         &\multicolumn{6}{|c|}{$a_{0} = 0.45$ km}                      \\
         $\Delta a$ (km)       &    0.2155    & 0.3306 &   3.5644  &  0.0570    &  0.1265 & 6.8571  \\
         $\Delta e$            &    0.4045    & 0.1576 &   10.1408 &  0.1394    &  0.1265 & 6.3158  \\
         $\Delta i (^{\circ})$ &    11.6476   & 0.1680 &   1.3433  &  6.5521    &  0.3784 & 2.6471   \\
         &\multicolumn{6}{|c|}{$a_{0} = 0.50$ km}               \\
         $\Delta a$ (km)       &    0.2047    & 0.1266 &   2.5043  &  0.0474    & 0.1265 &  6.0251 \\
         $\Delta e$            &    0.3004    & 0.1540 &   9.1139  &  0.0953    & 0.1265 &  5.6031  \\
         $\Delta i (^{\circ})$ &    16.3133   & 0.1308 &   1.9726  &  4.5566    & 0.1265 & 1.4076  \\
         &\multicolumn{6}{|c|}{$a_{0} = 1.00$ km}               \\
         $\Delta a$ (km)       &    0.0221    & 0.1472 &  2.0870 & 0.0075   & 0.1266 & 3.0316  \\
         $\Delta e$            &    0.0374    & 0.1265 &  2.5760 & 0.0125   & 0.1265 & 1.5352  \\
         $\Delta i (^{\circ})$ &    3.2891    & 0.1290 &  0.1386 & 0.3560   & 0.1265 & 1.5047  \\
         \hline
      \end{tabular}}
      \end{center}
      \end{minipage}
      \end{table}

   \section{Study of the orbital stability}\label{sec05_stability_region}
   
\begin{figure}[!ht]
         \centering{\includegraphics[width=0.85\linewidth]{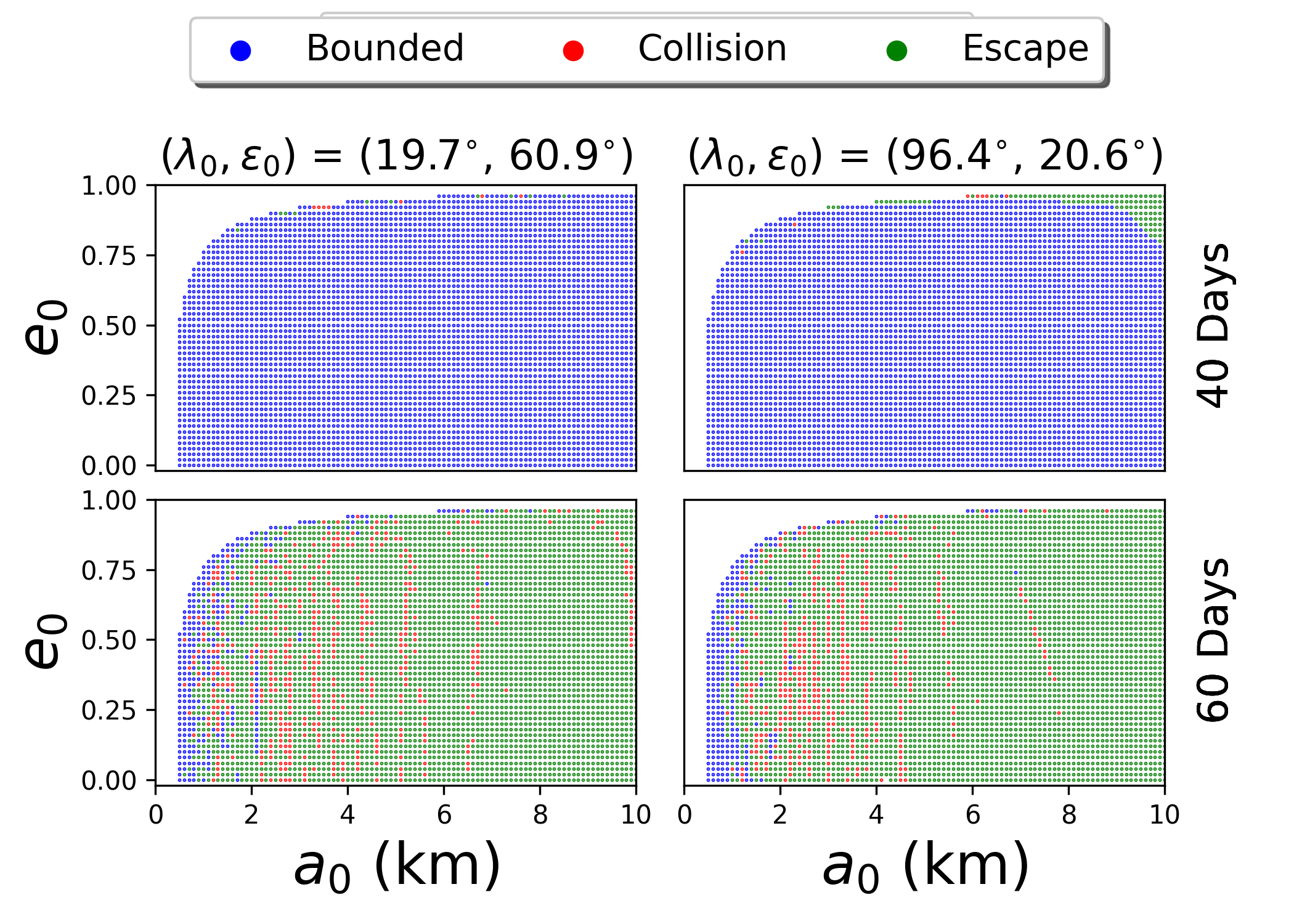}}
         \caption{Characterization of the orbits around (99942) Apophis for 40 days (Top) and 60 days (Bottom) time span starting from March 1, 2029.} \label{fig05_type_orbits}
      \end{figure}

      In this section, we carry out a qualitative analysis of the orbital stability of a spacecraft around Apophis before and after the close encounter with the Earth. We test the two extreme cases of initial conditions of Apophis spin orientation described in Sect.2, that is to say ($\lambda_{0}$, $\varepsilon_{0}$) = (19.7$^{\circ}$, 60.9$^{\circ}$) and (96.4$^{\circ}$, 20.6$^{\circ}$), referred hereafter as Spin-1 and Spin-2. We consider the full set of perturbations on the spacecraft, mentioned in Sect. \ref{sec03_Dynamical_model}. We use eq. \ref{motion1} to describe the 60-days motion of the spacecraft around Apophis. The initial conditions of the planets are generated by HORIZONS and set on March 1, 2029. Thus, our 60 days time span covers 43 days before the close encounter and 16 days after. In Fig. \ref{fig05_type_orbits}, we show the final states of the orbits integrated for 40-days (top panels) and 60-days (bottom panels). An orbit is considered to characterize an escape from the asteroid when the distance from its centre becomes 3 times larger than the Apophis Hill sphere. It is considered to characterize a collision with the central body when the particle crosses a 3D ellipsoid of radius 0.235 $\times$ 0.189 $\times$ 0.176 km. We notice that the large majority of the orbits ($\sim$ 95\%) collide or escape from the system just after the close encounter with our planet, whereas the totality of orbits are bounded before. This confirms the conclusion of \citet{aljbaae_2021}. We also remark that the initial Apophis spin orientation slightly affects the distribution of the colliding and escaping orbits. This demonstrates that it has to be taken into account for the computations. 
   
\begin{figure}[!ht]
         \includegraphics[width=\linewidth]{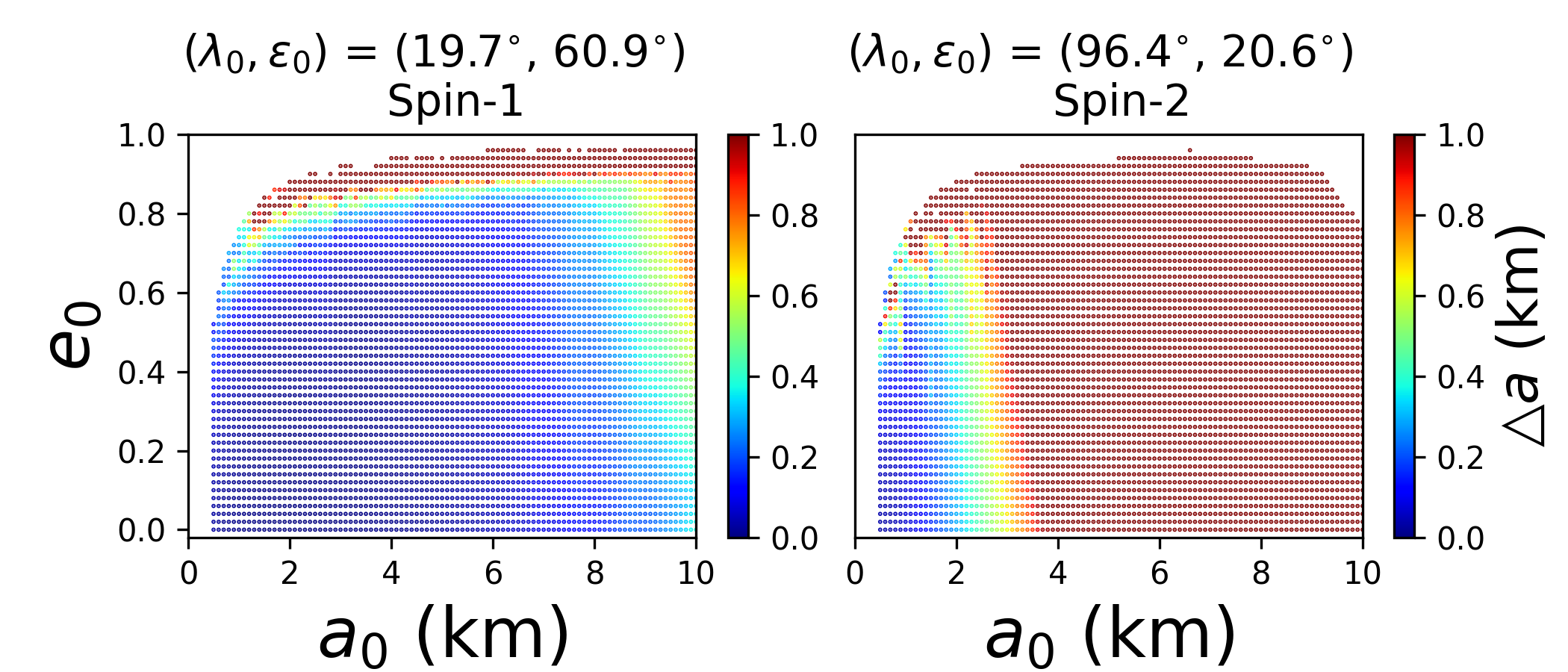}
         \caption{Variation maps of the semi-major axis coming from the ensemble perturbations on the real system of Apophis before the close approach with the Earth.} \label{fig06_stable_orbits}
      \end{figure}

\begin{figure}[!ht]
         \includegraphics[width=0.49\linewidth]{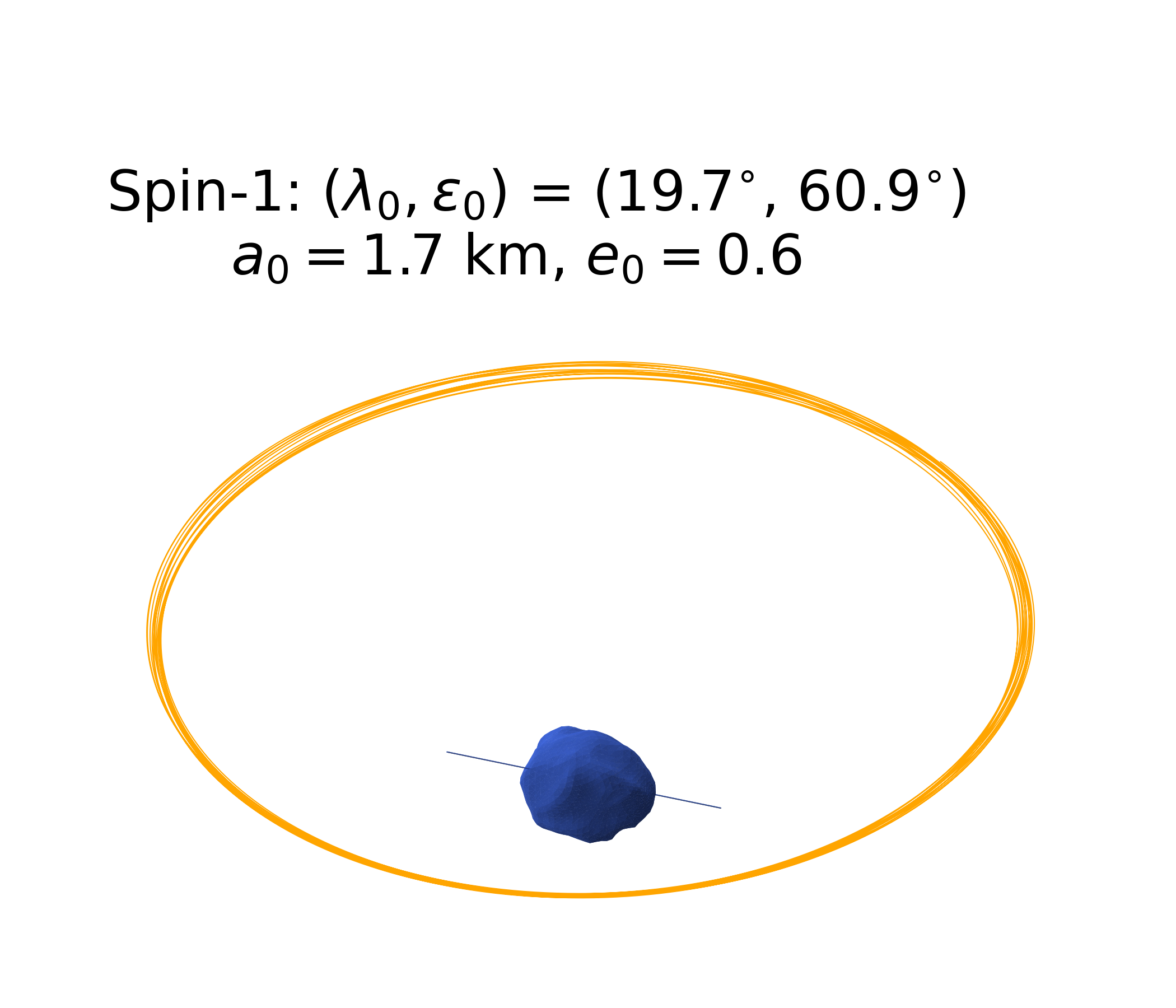}
         \includegraphics[width=0.49\linewidth]{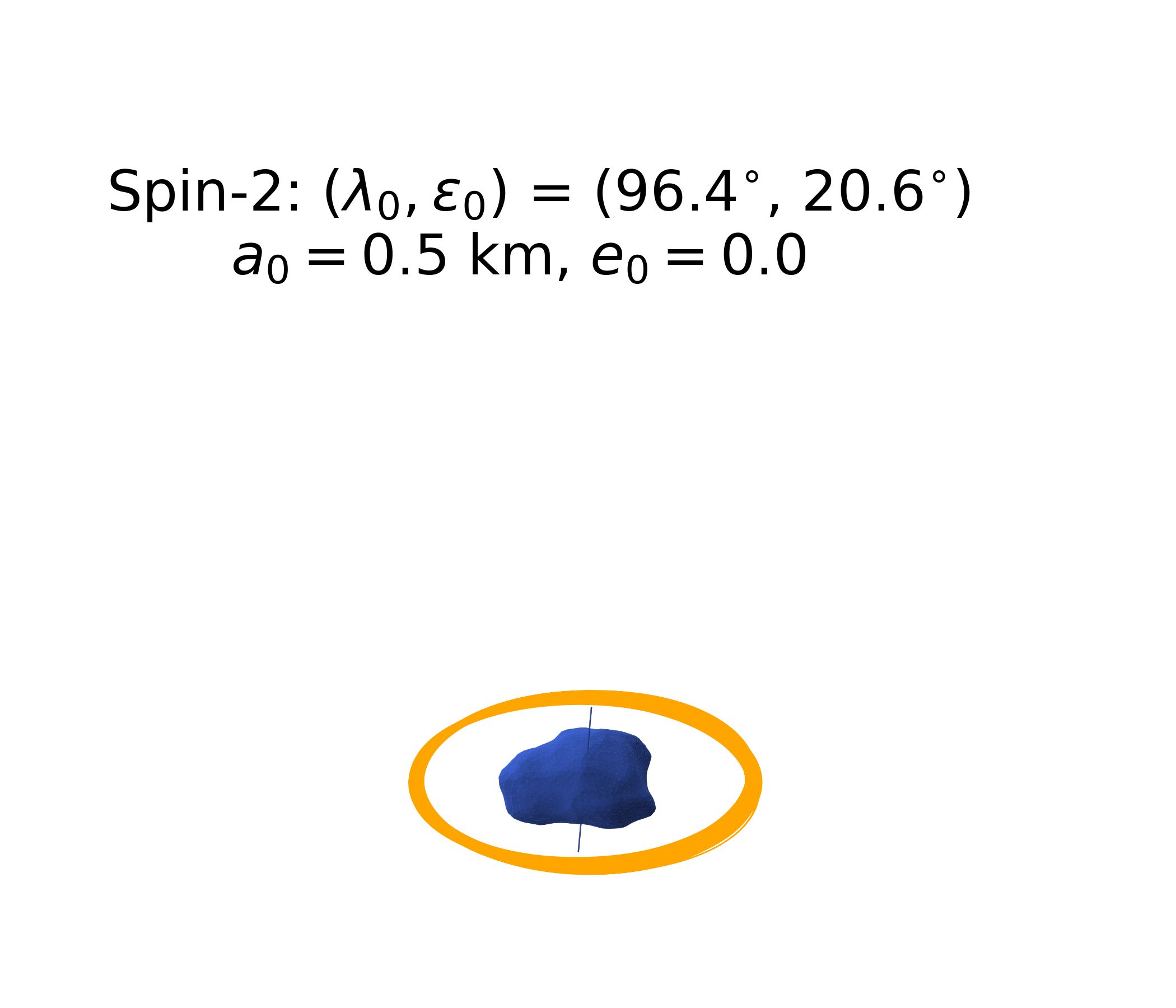}\\
         \caption{Less perturbed orbits around Apophis before the close approach with our planet.} \label{fig07_orbits}
      \end{figure}

      In Fig. \ref{fig06_stable_orbits}, the level of perturbation after a 40-day integration is characterized by the peak-to-peak amplitude ($\Delta a$) of the variations of the semi-major axis. Moreover, we observe that the less perturbed region generated by the case with Spin-1 initial conditions is by far more extended than the one generated by Spin-2. This should come from the irregularity of the projection of the shape of the asteroid on the spacecraft orbital plane. The less perturbed orbit in each case is shown in Fig. \ref{fig07_orbits}, which corresponds to $\Delta a = \sim 2$ m for Spin-1 and $\Delta a = \sim 35$ m for Spin-2. \\
   
     As noted earlier in this section, the large majority of orbits, integrated for 60-days time interval, collide or escape because of the close encounter between Apophis and our planet. Indeed, the heavy perturbed system of a particle surrounding Apophis under the influence of the Earth tends to produce by far more chaotic than regular orbits after the encounter. In order to investigate the phase space structure of the orbits, we apply the following three different methods  of analysis. 
     \subsection{Use of the MEGNO algorithm}
\begin{figure}[!ht]
            \includegraphics[width=\linewidth]{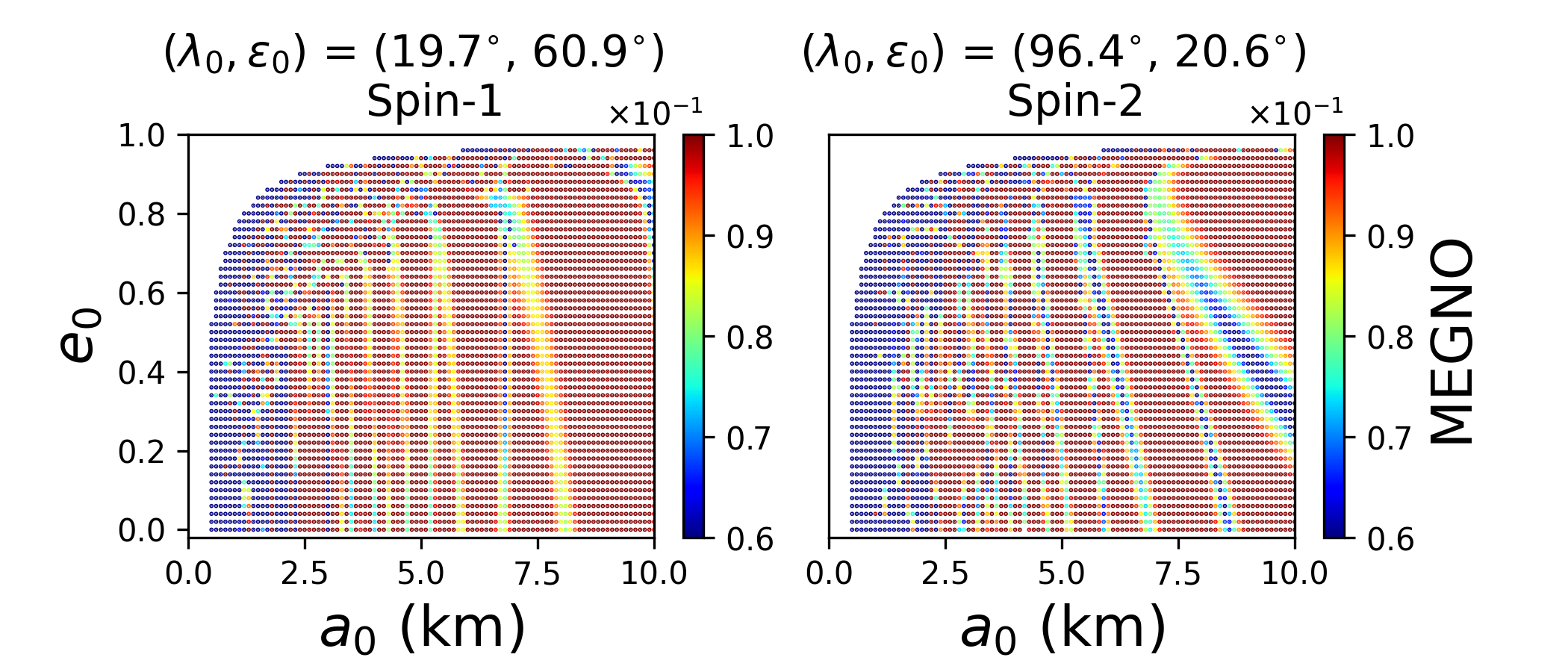}
            \caption{MEGNO dynamical maps for the spacecraft orbits around the Apophis system.} \label{fig08_megno}
        \end{figure}

        In this subsection, we apply the Mean Exponential Growth factor of Nearby Orbits developed by \citet{cincotta_2000}. In this method, a global dynamics insight is obtained by calculating the average of the relative divergence of the orbit using the following expression \citep{mestre_2011}
        \begin{eqnarray}\label{pi_method}
          \text{MEGNO} = \frac{2}{T}\sum_{k=1}^{T}k\ln \bigg(\frac{\delta(k)}{\delta(k-1)}\bigg)
        \end{eqnarray}
        \noindent where, $\delta(k)$ represents the deviation vector in the phase space, and $T$ is the time of integration. Our results are illustrated in Fig. \ref{fig08_megno}. The larger MEGNO values correspond to a higher degree of chaos and a higher chance of instability. We notice that there are some isolated regions (in dark blue color) where the spacecraft can maintain quasi-periodic orbits. Some straight lines appear in the map, which could indicate some type of resonance in the system. However, studying these resonances is out of the scope of this work. 
        
     \subsection{Perturbation Map of type $II$ (PMap)}
\begin{figure}[!ht]
            \includegraphics[width=\linewidth]{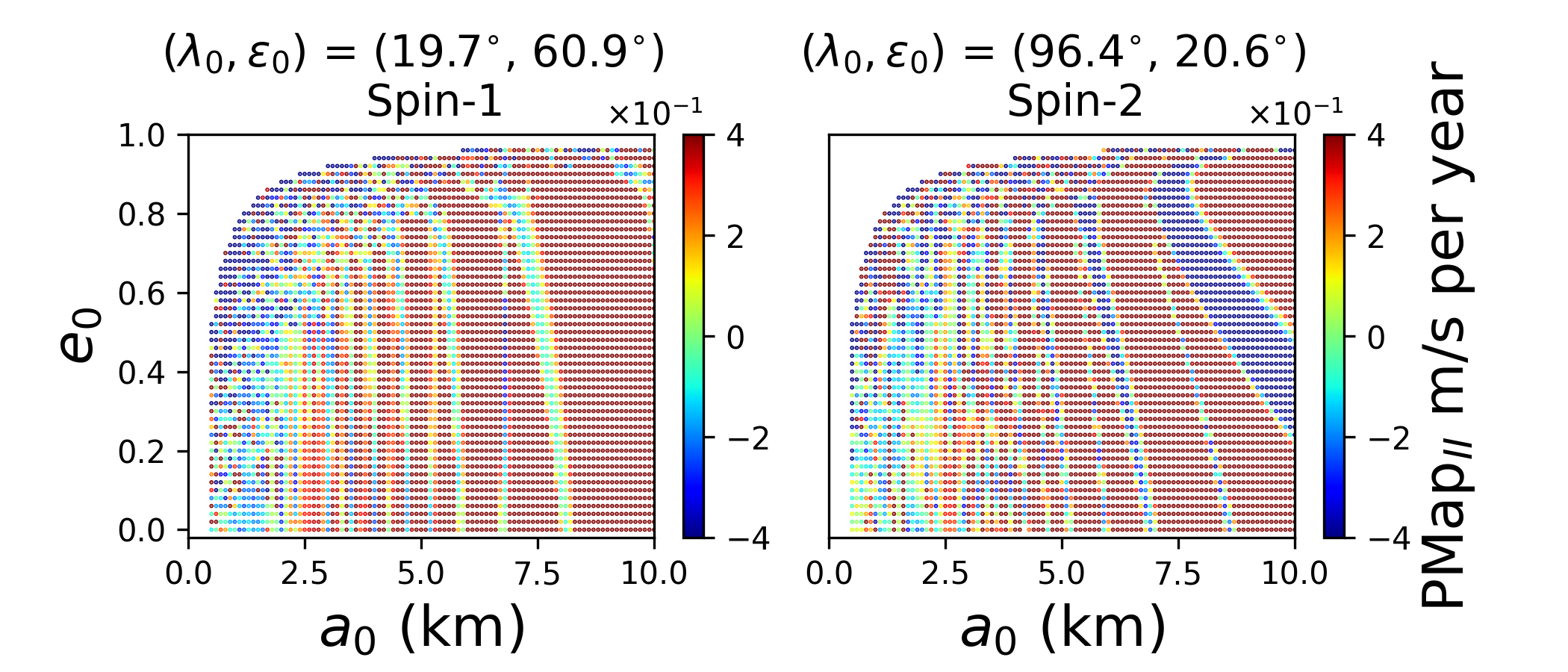}
            \caption{Perturbation maps of type II (PMap) for a spacecraft orbiting around Apophis.} \label{fig09_pi2}
         \end{figure}

        In this method, we calculate the Perturbation Maps of type II, as presented in \citet{sanchez_2017, sanchez_2019} and \citet{sanchez_2020}. The perturbations of energy undergone by the spacecraft are measured according to the following expression.
         \begin{eqnarray}\label{pi_method}
            \text{PI}_{ii} &=& \frac{1}{T} \int_{0}^{T} \langle \boldsymbol{a}, \frac{\boldsymbol{v}}{|\boldsymbol{v}|}\rangle dt,
         \end{eqnarray}
         \noindent where, $\boldsymbol{a}$ is the acceleration due to the whole perturbations of the orbital motion, $\boldsymbol{v}$ is the velocity of the spacecraft, $T$ is the final time of the numerical integration. In this approach, the value of $\text{PI}_{ii}$  gives a good indication of the variation of energy caused by the perturbations. For instance, the blue zone in Fig. \ref{fig09_pi2} corresponds to a negative value of the integral, which indicates a loss of energy and, as a consequence, a decreasing semi-major axis, which could lead to a collision with the asteroid. However, not all collisional orbits have a negative value of this integral. If the orbit collides after about ten days of integration, the integral value is positive, as we will see later. We notice that the zone with negative values of $\text{PI}_{ii}$ is compatible with the smallest MEGNO values shown in Fig. \ref{fig08_megno}. In fact, MEGNO is a tool principally devoted to detect chaos, and after 10 days of integration, the trajectories may not present chaotic behaviour, even if they are highly disturbed as shown by the PMap algorithm. In other words, we can conclude that the PMap method can provide more information close to the central body than the MEGNO method.             
      \subsection{Time-Series prediction}
\begin{figure}[!ht]
            \includegraphics[width=0.48\linewidth]{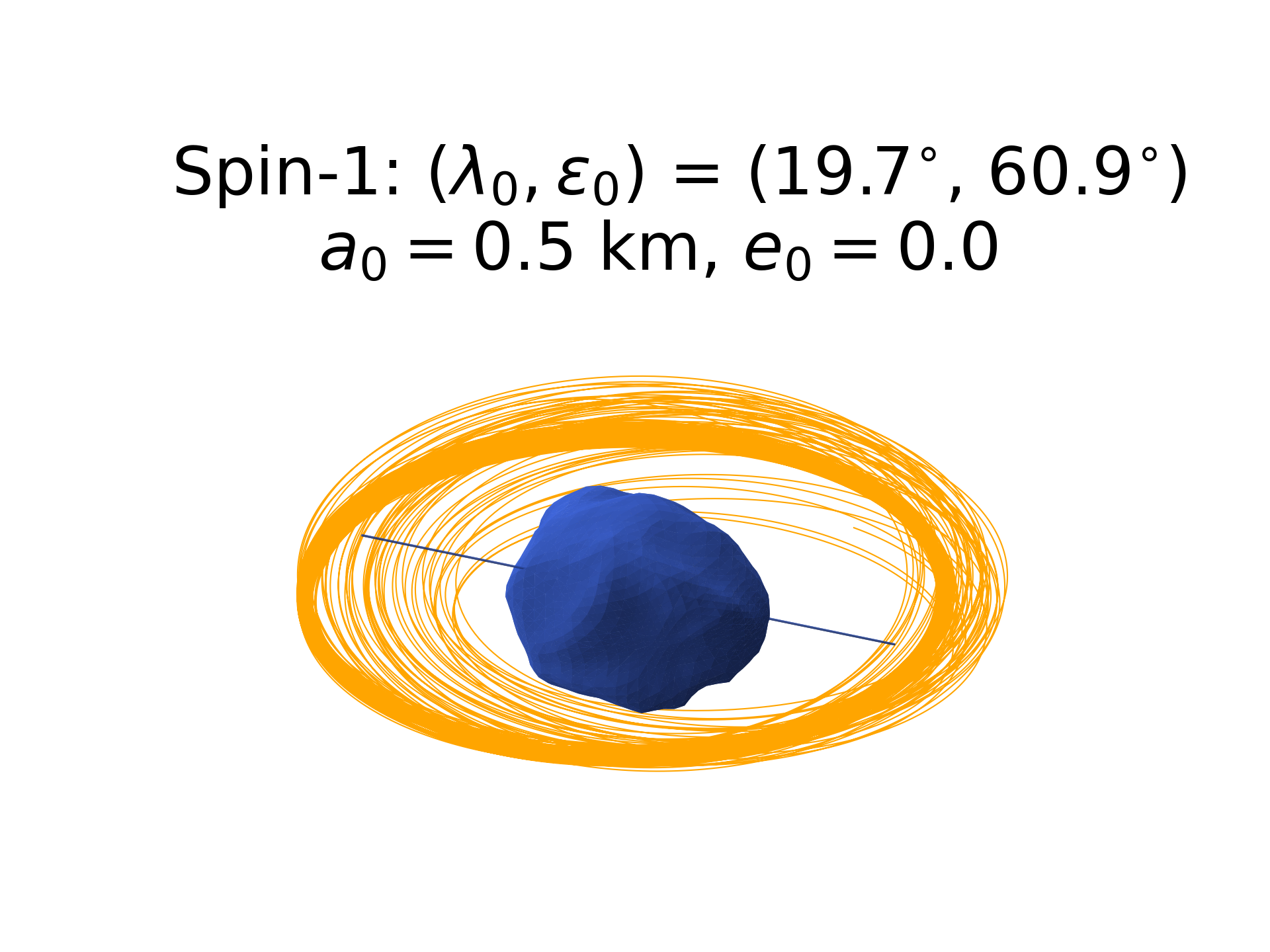}
            \includegraphics[width=0.48\linewidth]{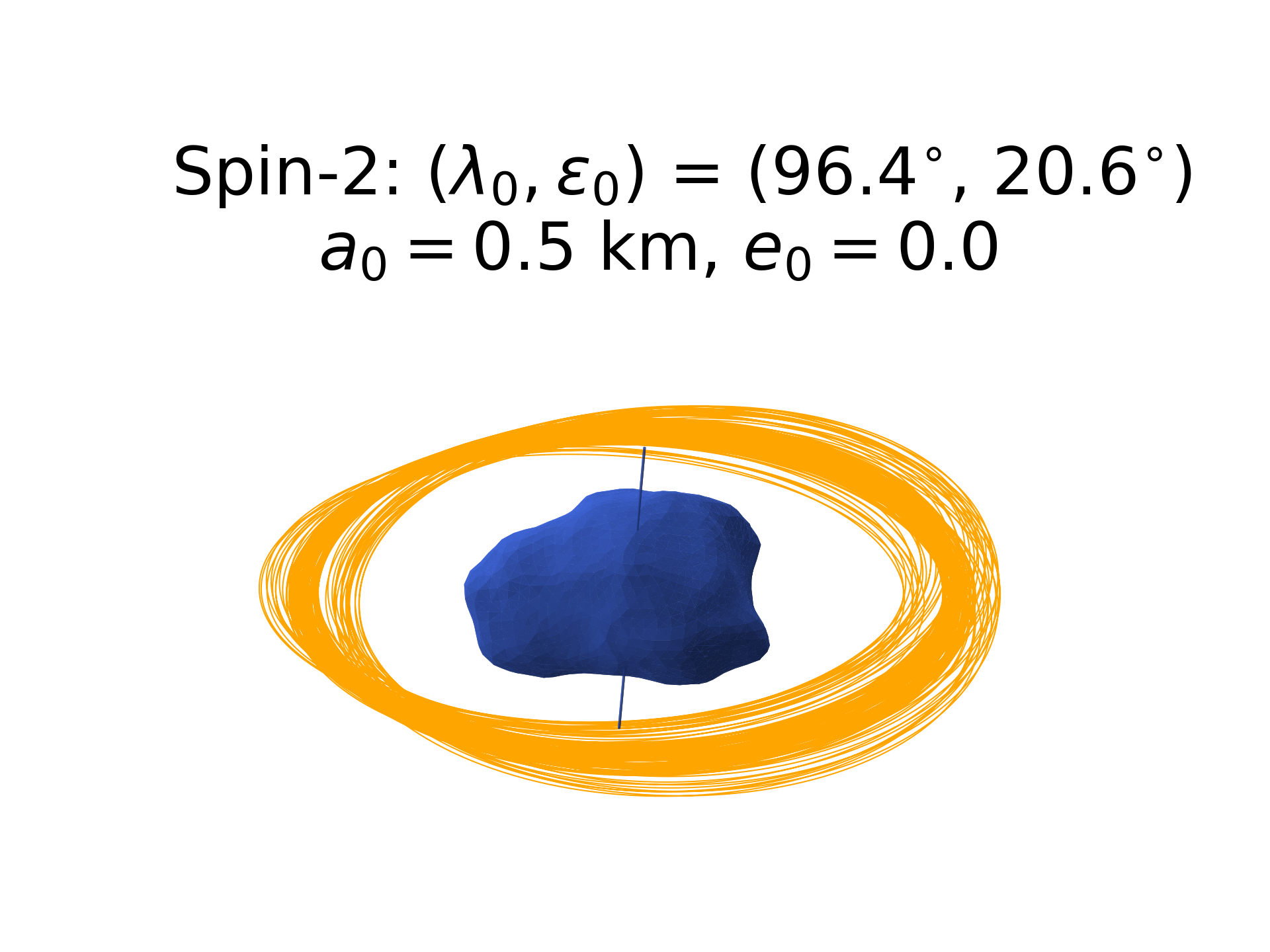}\\
            \includegraphics[width=0.48\linewidth]{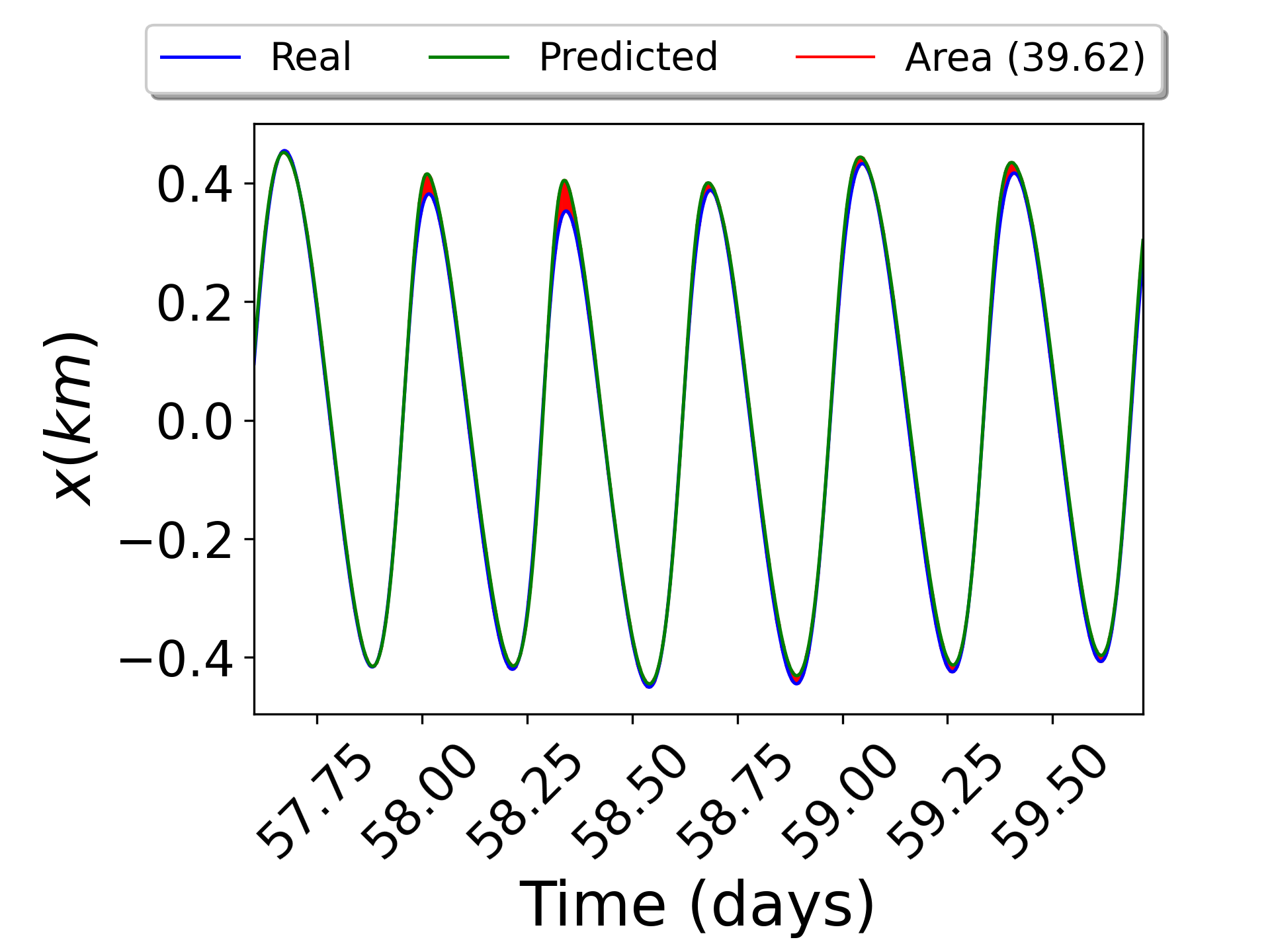}
            \includegraphics[width=0.48\linewidth]{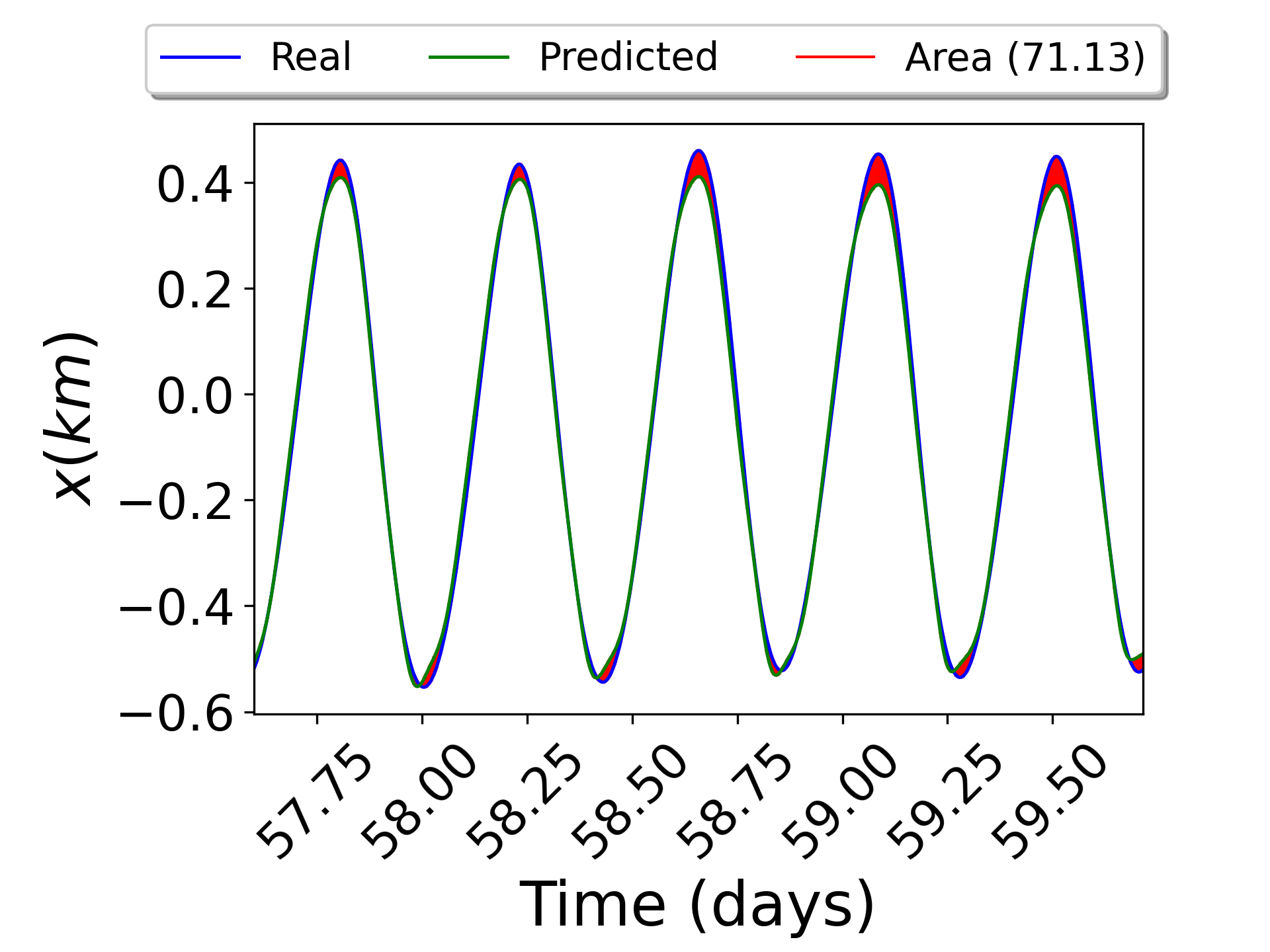}\\
            \includegraphics[width=0.48\linewidth]{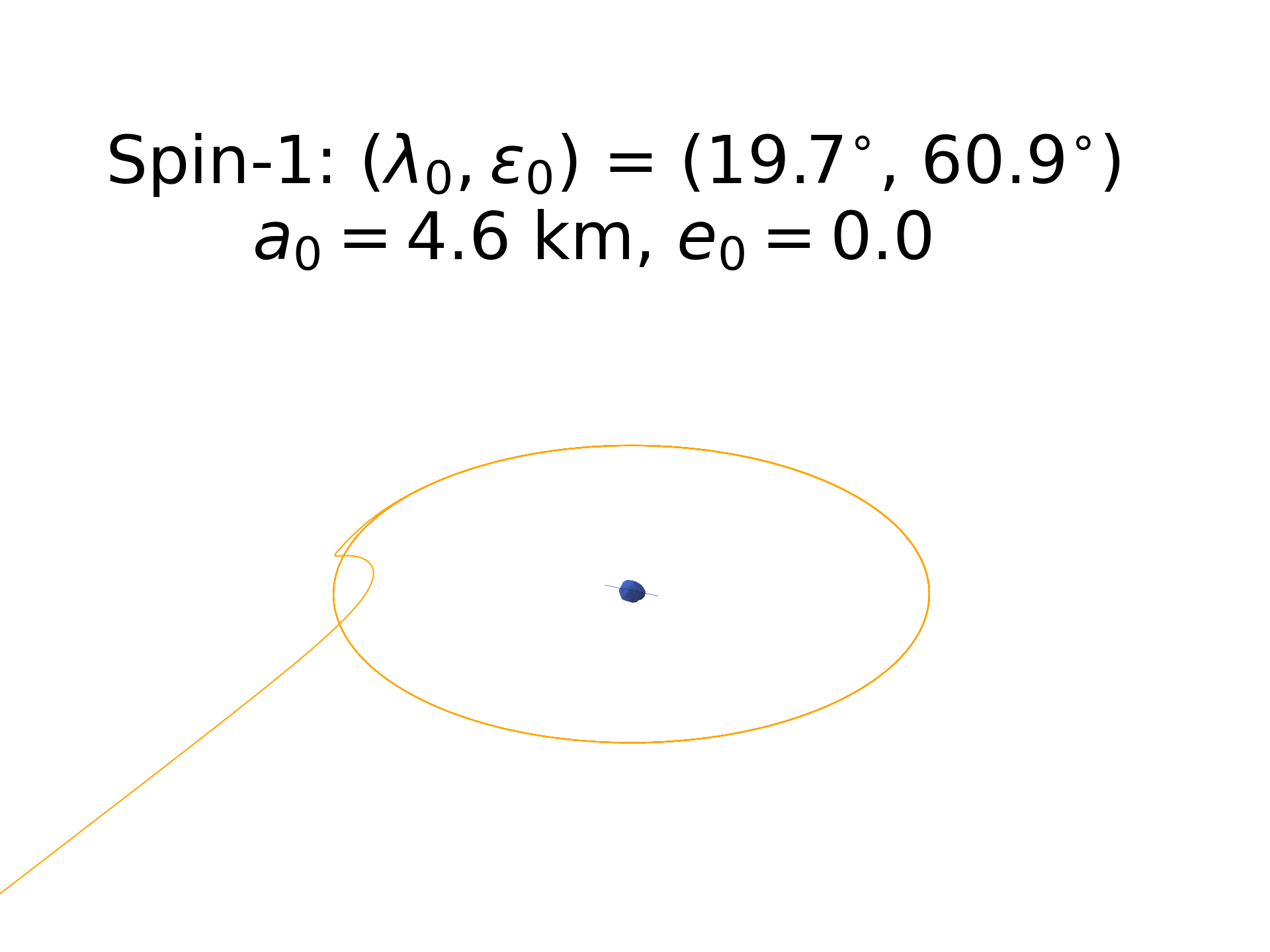}
            \includegraphics[width=0.48\linewidth]{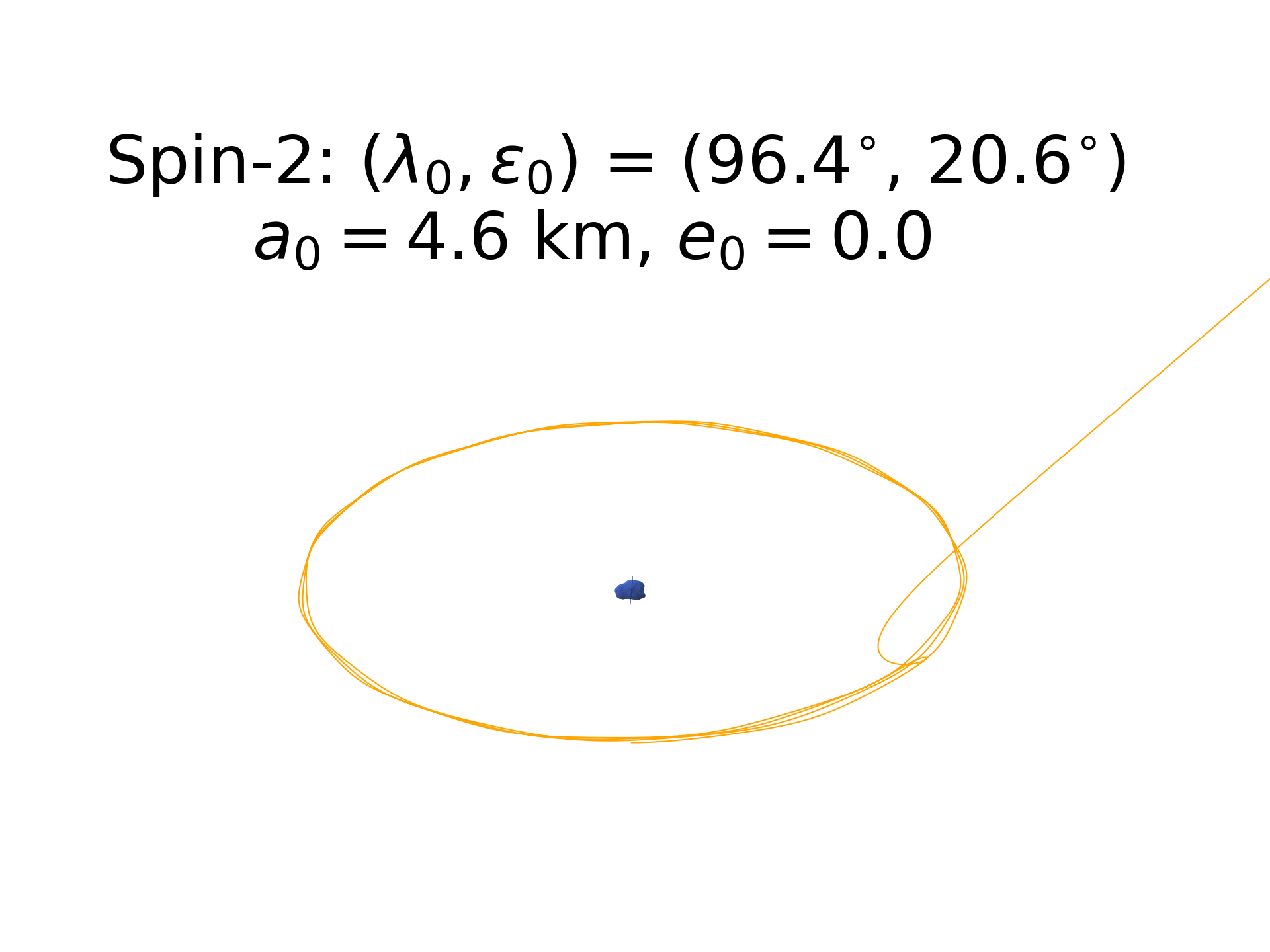}\\
            \includegraphics[width=0.48\linewidth]{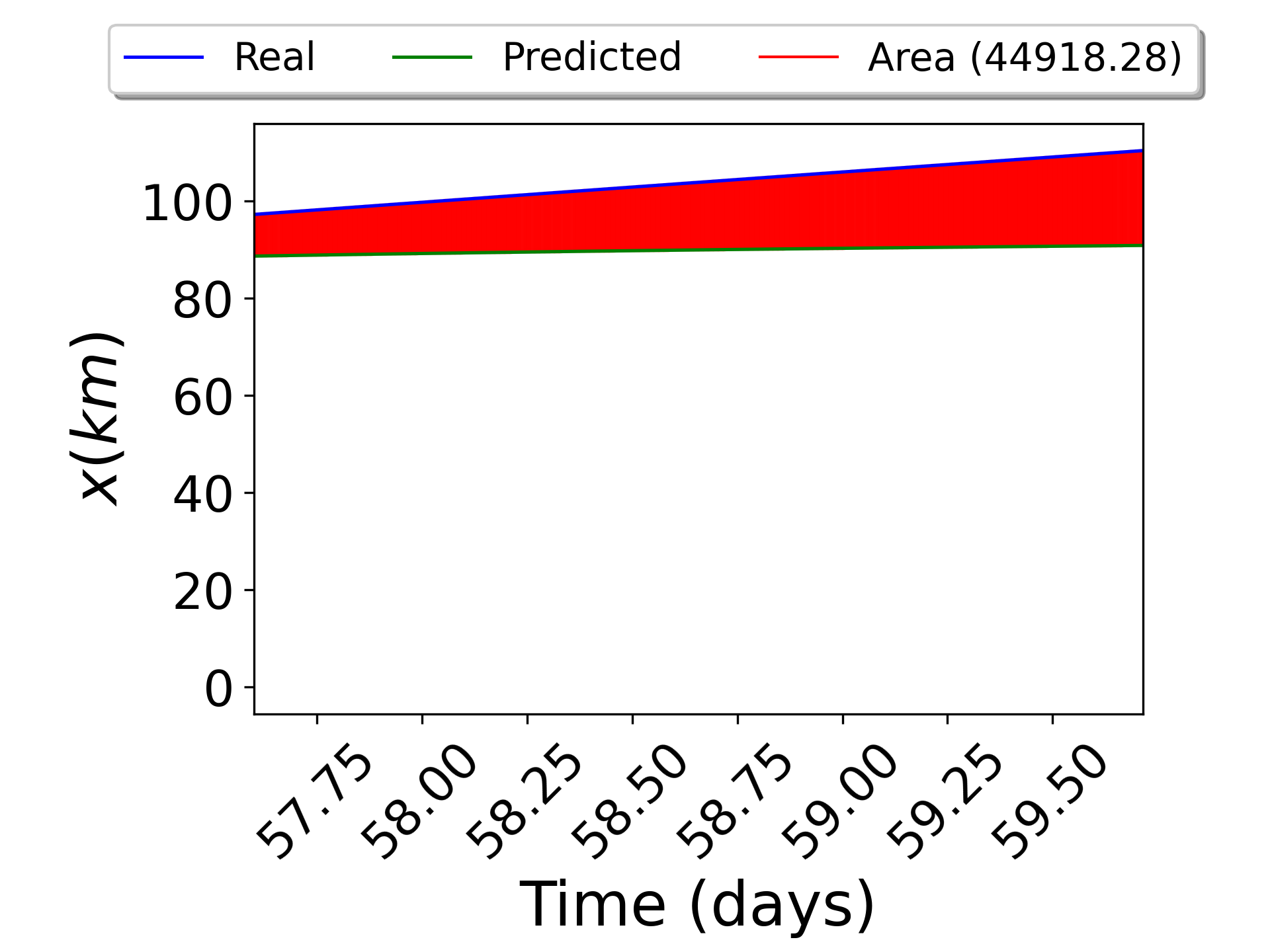}
            \includegraphics[width=0.48\linewidth]{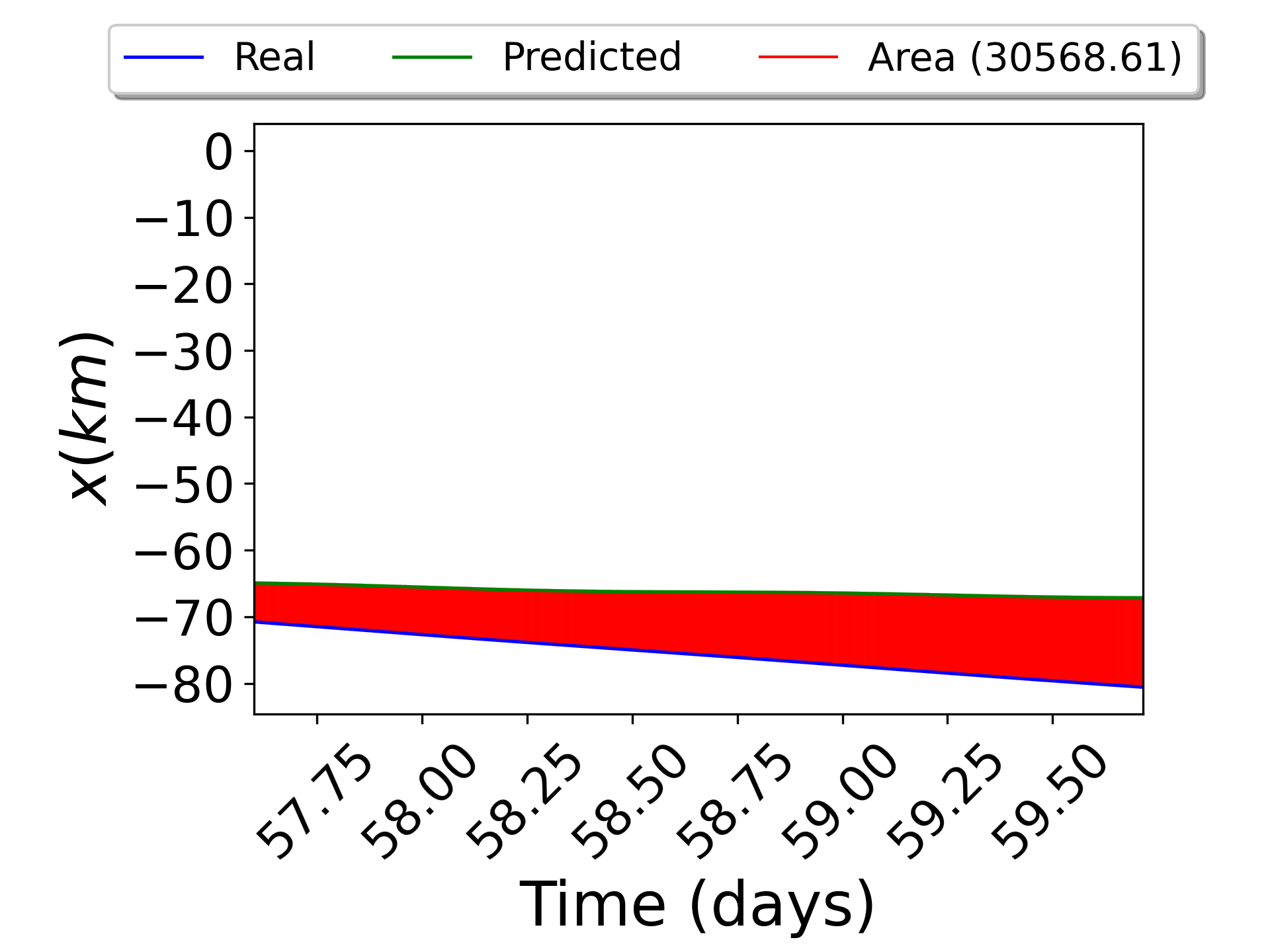}\\
            \caption{Example of a regular (Top panel) and an irregular orbit (Bottom panel) for a spacecraft orbiting around Apophis.} \label{fig10_time_series_example}
         \end{figure}

         Unlike the previous two methods, this one does not use the nearby orbits. It relies on computational intelligence in Time-Series forecasting to predict the behaviour of the distribution of orbit coordinates based solely on the past-patterns. We built our model using \textsc{python} language, \textsc{Keras} \citep{chollet_2015}, and \textsc{Tensorflow} frameworks \citep{abadi_2016}. The method consists in using a sequence of random variables to create a model fitted to historical data and to apply it to predict the future. The dataset for each orbit consists of 6 features (positions and velocity), recorded every 30 seconds. Each feature has values with varying ranges different from others. Thus, we normalize all the feature values between 0 and 1 before training a neural network. The first 90\% of the points in each orbit (54 days) are used to train the model and predict the position of the spacecraft during the last 6 days of the orbit. To optimize the performance, our model collects data for the first 12 hours (1440 observations), which are sampled every 2.5 minutes. The positions after 15 observations are used as a label. Our training is interrupted when the validation loss is no longer improving. More details about the method are presented in the Keras documentation pages available at \href{https://keras.io/}{https://keras.io/}. Then, we calculate the area between the predicted and real data ($\mathcal{A}$). The smaller the area, the more predictable the orbit, which makes the spacecraft mission much easier to be mapped and planned out. In Fig. \ref{fig10_time_series_example} we present an example of a bounded orbit and an orbit undergoing an escape after about 45 days, considering both the Spin-1 and Spin-2 conditions.   
         
\begin{figure}[!ht]
             \includegraphics[width=\linewidth]{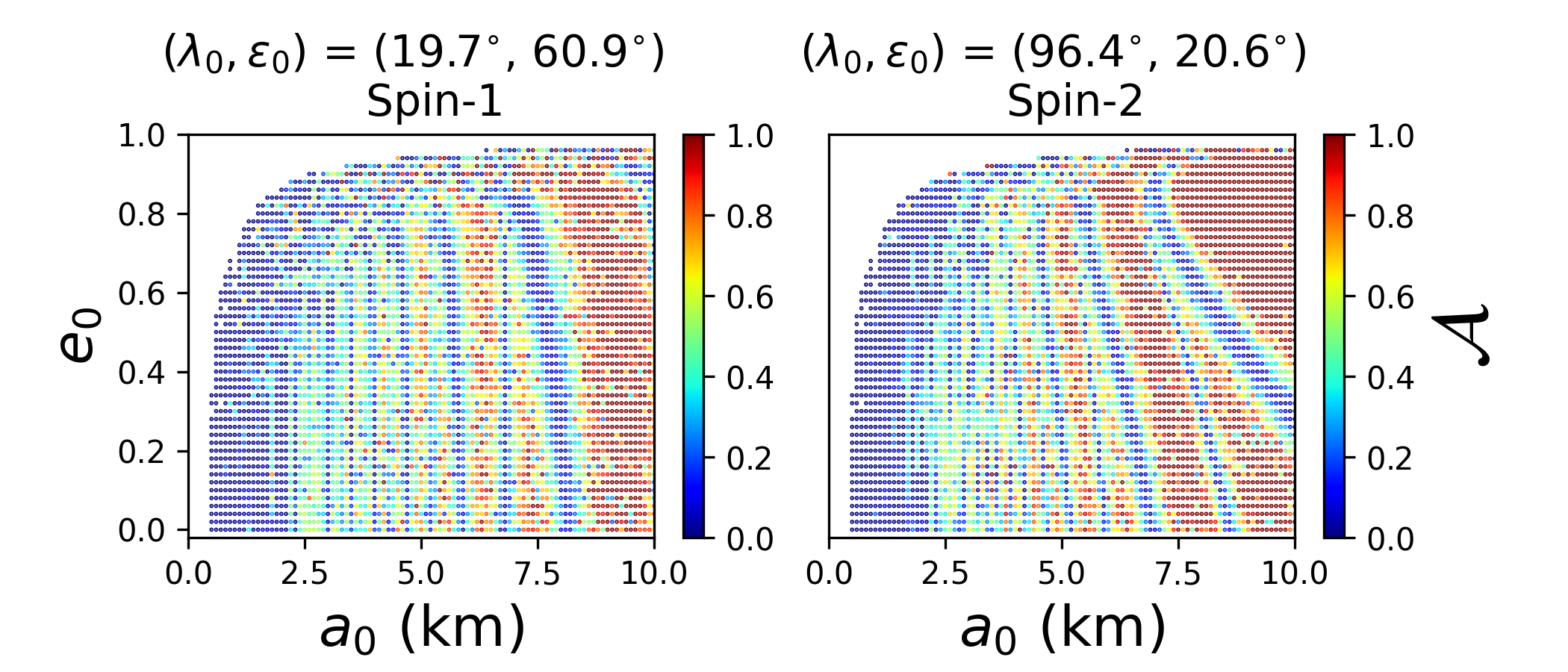}
             \caption{Forecasting maps using the Time-Series prediction for orbits around Apophis.} \label{fig10_time_series}
          \end{figure}

\begin{figure}[!ht]
             \includegraphics[width=0.48\linewidth]{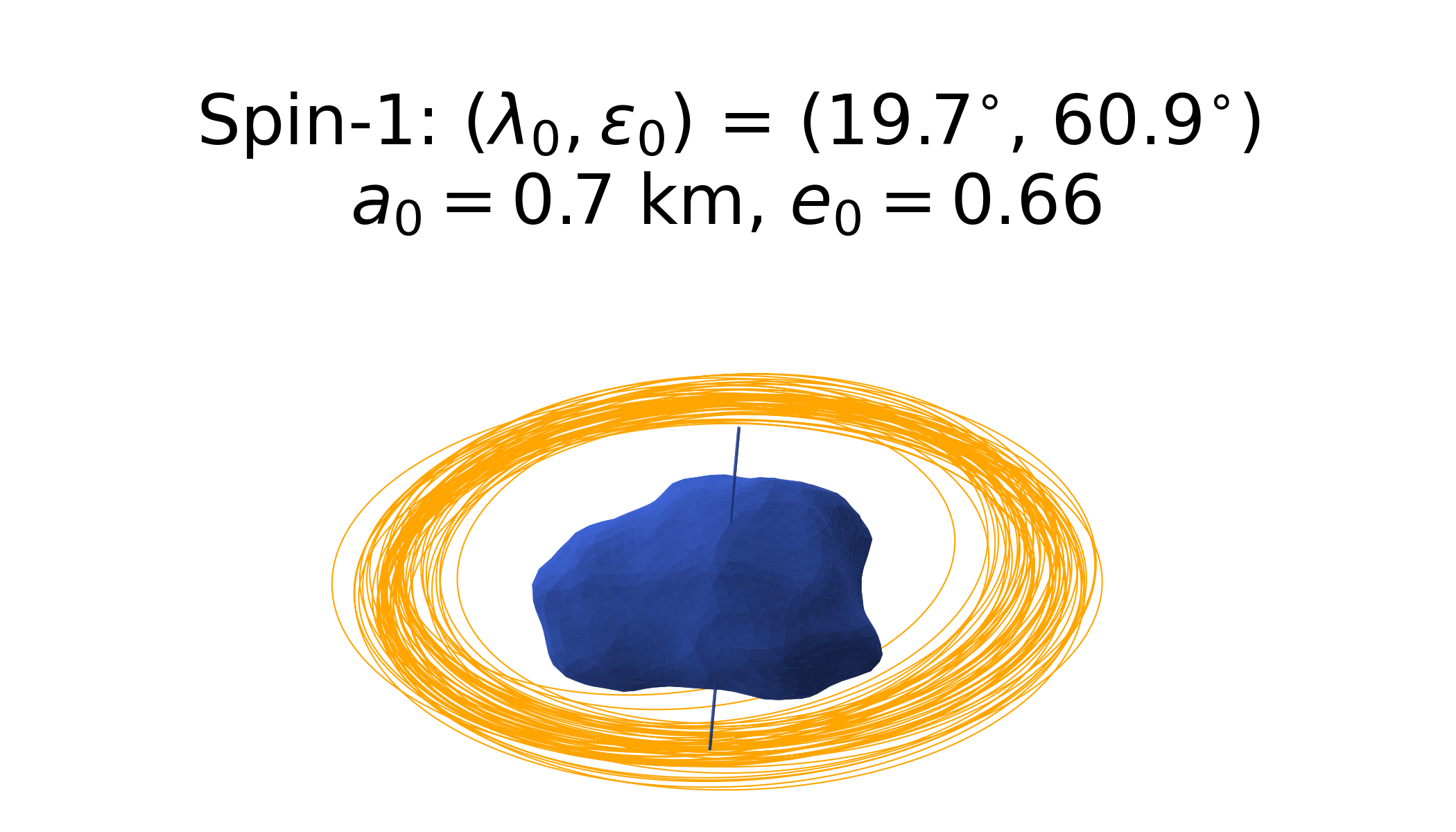}
             \includegraphics[width=0.48\linewidth]{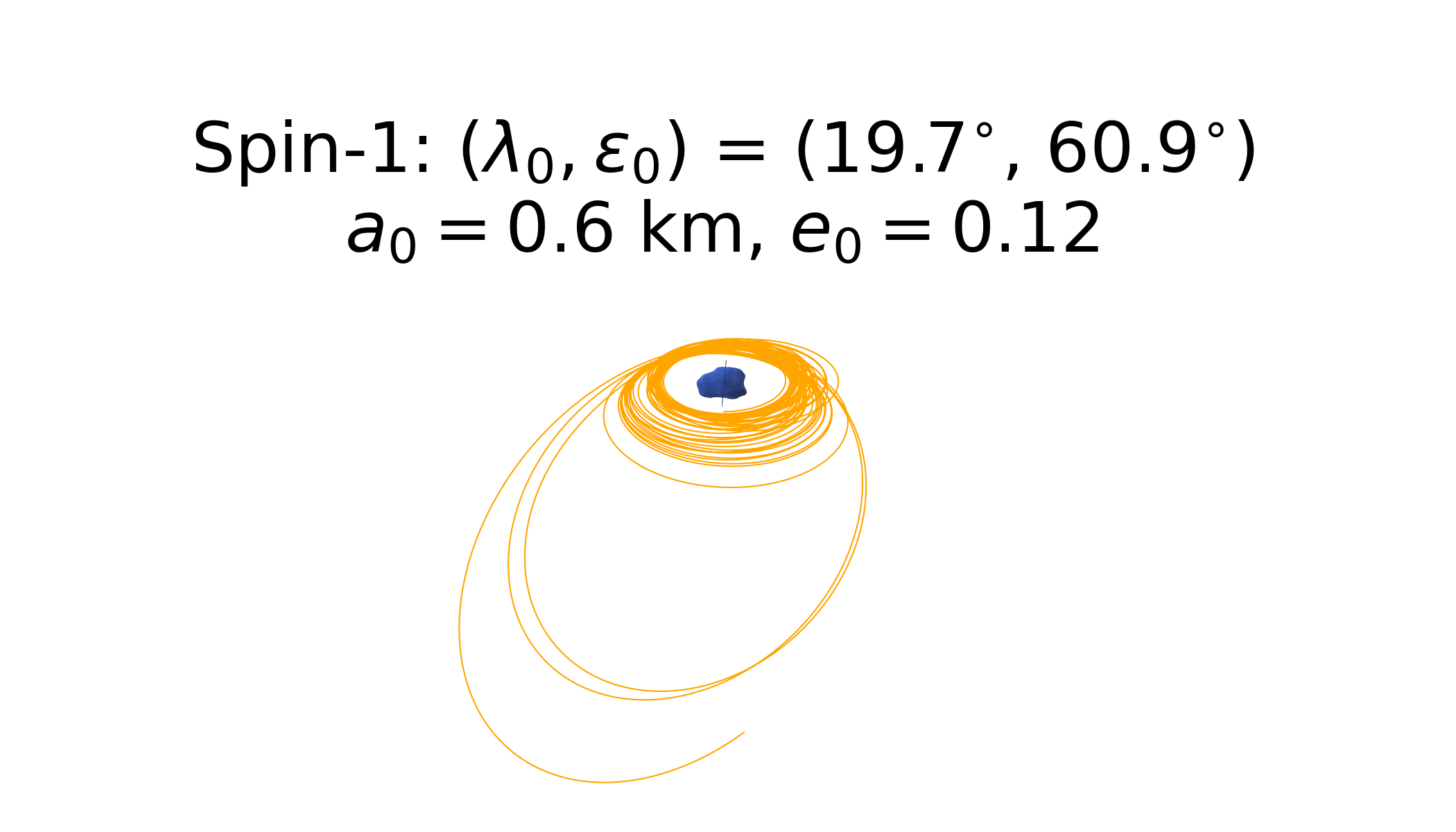}\\
             \includegraphics[width=0.48\linewidth]{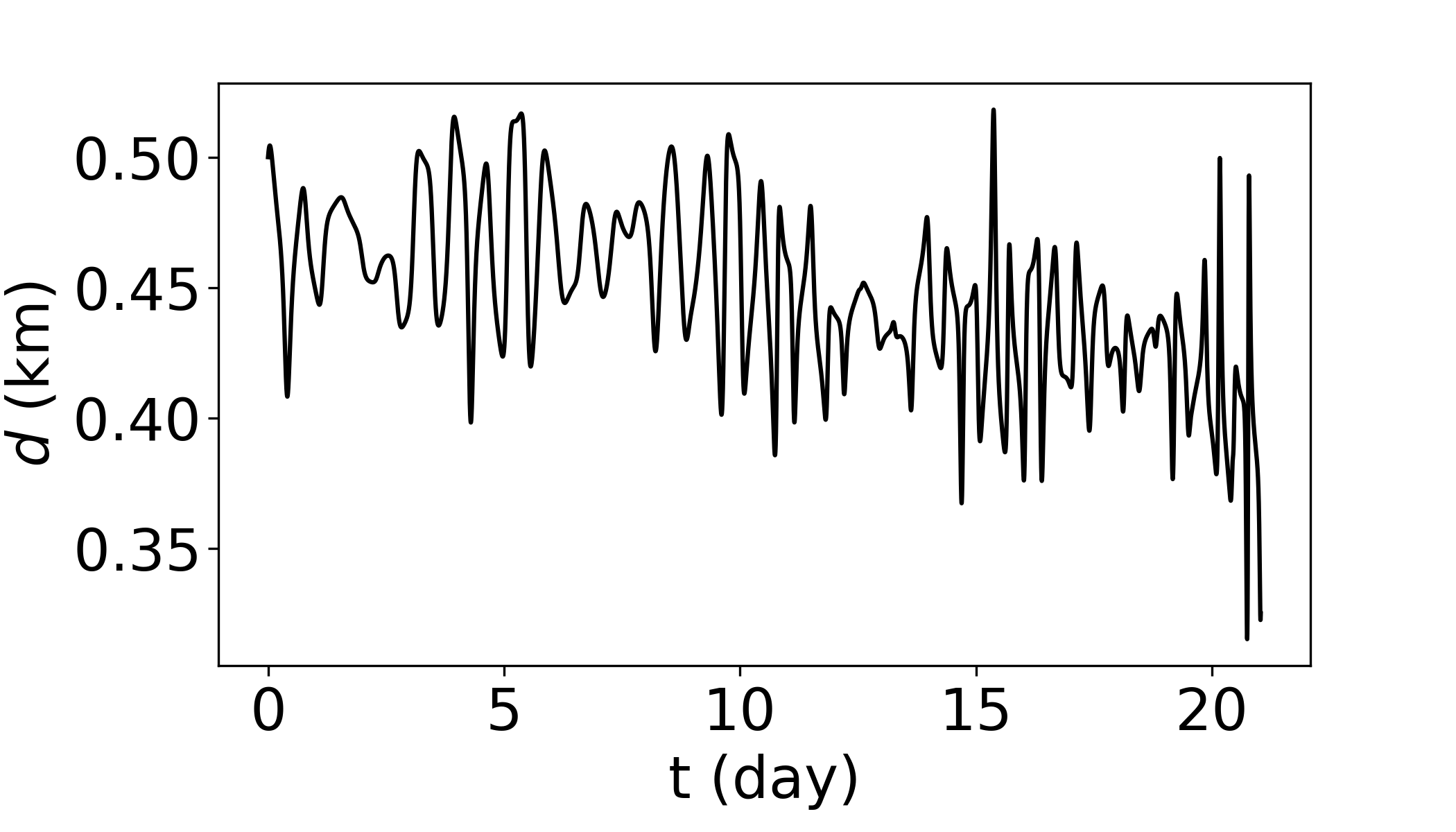}
             \includegraphics[width=0.48\linewidth]{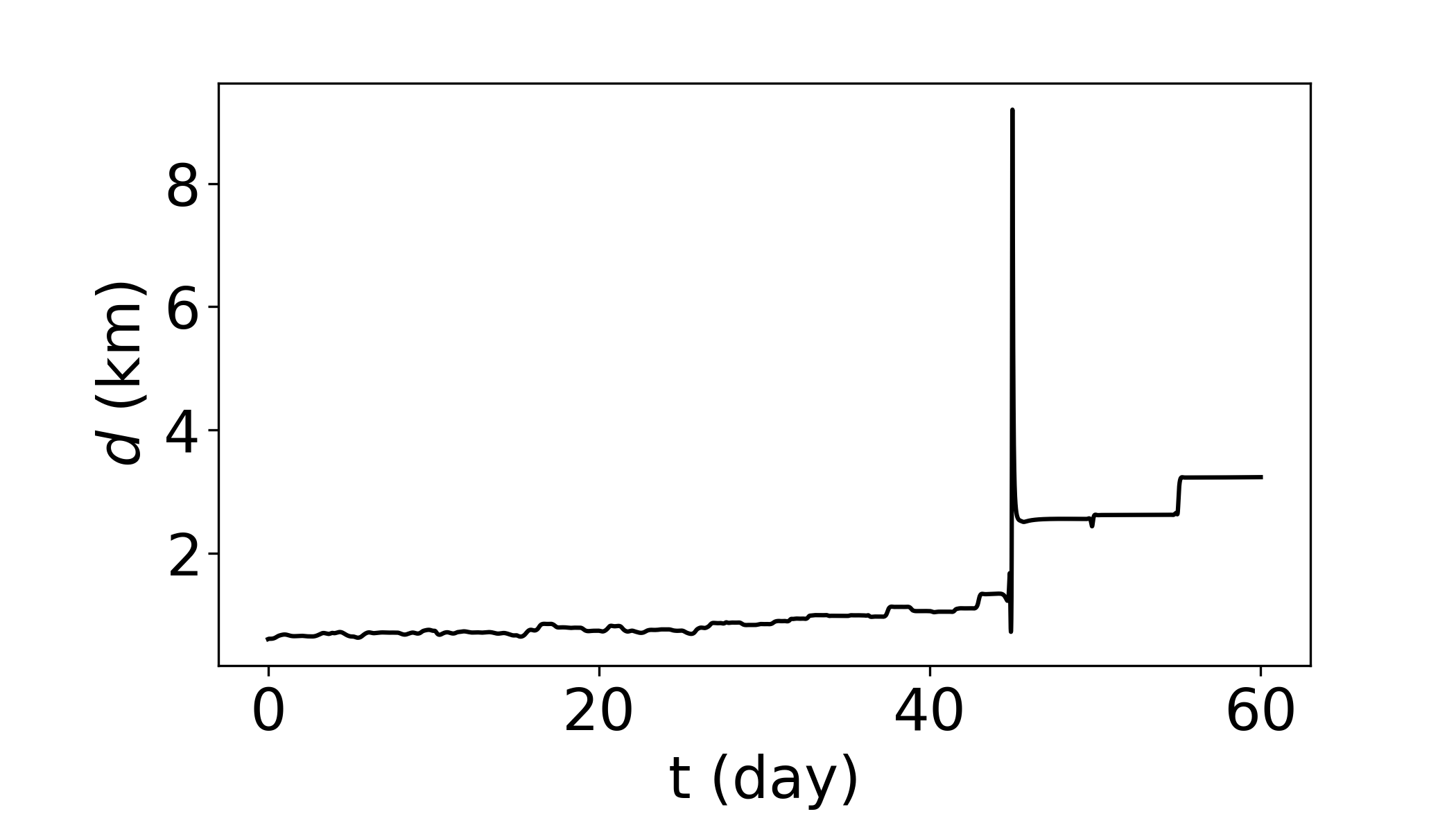}\\
             \caption{example of two neighbours orbits in the system of Apophis.} \label{fig_14_example_orbits}
          \end{figure}

          The Forecasting map using the area $\mathcal{A}$ normalized between 0 and 1 is presented in Fig. \ref{fig10_time_series}. From Figs. \ref{fig08_megno}, \ref{fig09_pi2}, and \ref{fig10_time_series} we notice that the three methods investigated here are conceptually compatible. Although the results are quite similar after $a_{0}=$ 5.0 km, it seems that PMap shows more details for the orbits closer to the asteroid. For instance, taking the two neighbouring orbits ($a_0=0.5$, $e_0=0.12$) and ($a_0=0.6$, $e_0=0.12$), we notice that the first orbit collides with the asteroid after 21 days, while the second one survives the 60-days integration (Fig. \ref{fig_14_example_orbits}). These orbits are stated in the PMap in a different category taking the values of 0.8 m/s and -0.45 m/s per year, respectively. However, they are represented by close values in the MEGNO and Time-Series prediction maps.
          
          \subsection{Comparison of the three methods}

\begin{figure}[!ht]
             \includegraphics[width=0.48\linewidth]{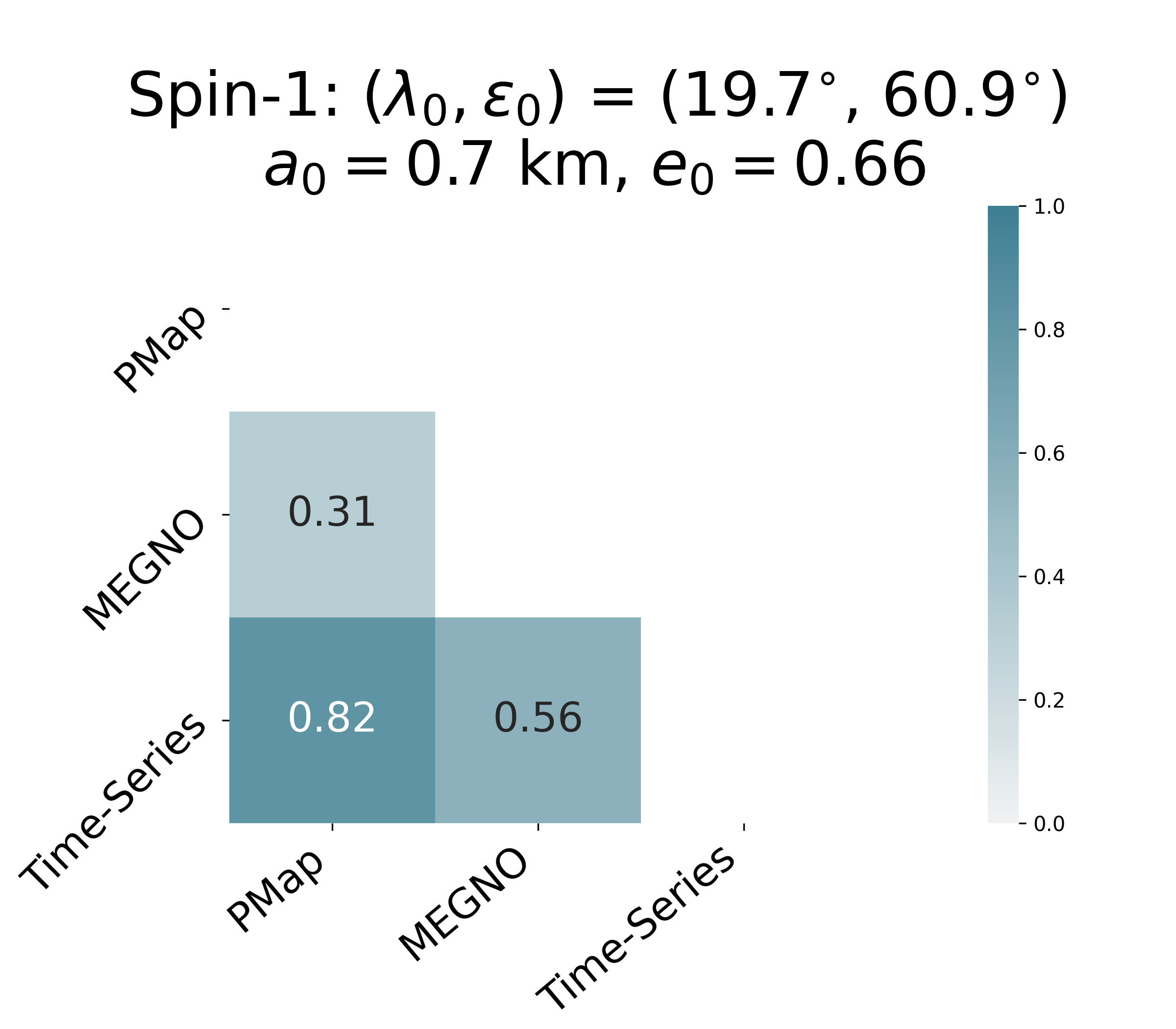}
             \includegraphics[width=0.48\linewidth]{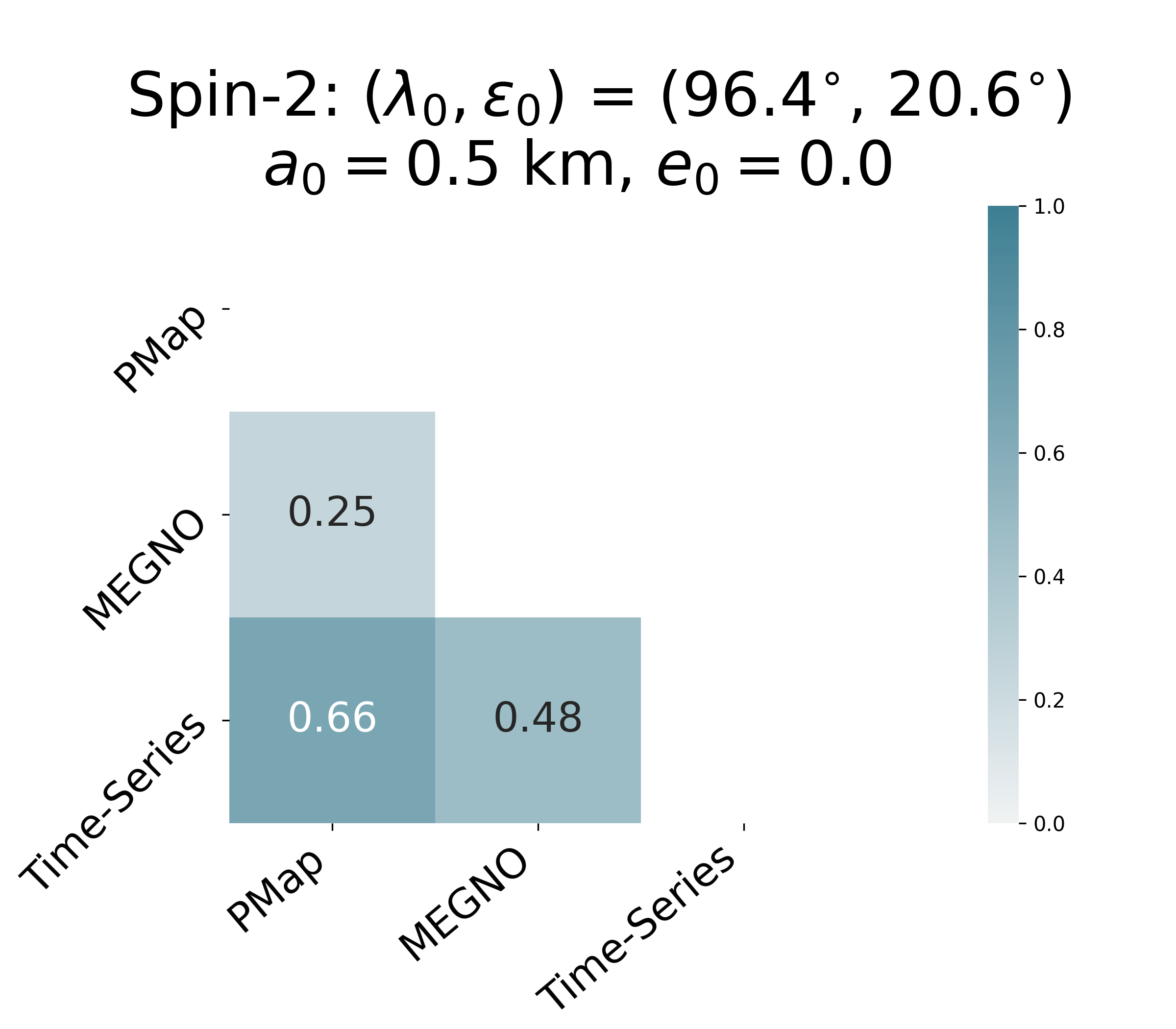}\\
             \caption{Correlation matrix for the three methods used to investigate the phase space structure associated to Apophis.} \label{fig13_correlation}
          \end{figure}

          Finally, we evaluate the coherence between the three methods presented above, using the Pearson correlation coefficient \citep{pearson_1895}, which measures a linear relationship between two given variables, denoted by the standard formula presented in \citet{carruba_2021}, where the authors identified the similarity between four chaos indicators: the Fast Lyapunov exponents \citep{froeschle_2000}, MEGNO, the frequency analysis method \citep{laskar_1990}, and the auto-correlation function \citep{carruba_2021}. The Pearson correlation coefficient always ranges from -1.0 (anti-correlation) +1.0 (correlation). As the value is close to 0 there is a dependence of the variables. Our results are presented in Fig. \ref{fig13_correlation}. We found that the PMap and the Time-Series methods are highly correlated.

    \section{Conclusion}
       In this work, we investigated the dynamics of a spacecraft orbiting around the asteroid (99942) Apophis during its 2029 close encounter with the Earth,  considering the dependence on the initial conditions of the rotation of the target. We used the dynamical model developed in our previous work \citep{aljbaae_2021}, for which we represented the gravitational field of the asteroid by a cloud of 3996 point masses system distributed inside a polyhedral shape derived from \citet{brozovic_2018}. We applied the method of \citet{souchay_2018} to determine the changes of Apophis' spin state due to the terrestrial torques during the close encounter. In a first step, we explored the impact of this phenomenon on the dynamics of a spacecraft orbiting around the asteroid, considering two cases of initial spin orientation, corresponding respectively to the minimum and maximum values of the spin variations during the encounter.
       
       We showed that this orientation can influence significantly the behaviour of the orbital motion of the spacecraft. In a second step we carried out a 60-days integration ranging 43 days before and 16 days after the encounter and studied the dependence of the stabilty of orbits with respect to the initial value of the semi-major axis. We found that the very large majority of cases, the spaceraft undergoes a collision or escape due to the perturbation caused by the close encounter, whereas it shows in all cases a very stable orbit before. Then we applied three different methods i.e. MEGNO, PMap and Time Series Forecasting to charcaterize in a deeper way the degree of stability or chaoticity of the orbits. The Time-Series Forecasting is used do classify orbits based on a relationship between the difficulty in the prediction and the stability. Using this method, we isolated the most predictable orbts that could be a stable ones in the system. A good correlation was found between this approach and MEGNO \citep{cincotta_2000} or the Perturbation Map of type $II$ (PMap) \citep{sanchez_2017, sanchez_2019}. However, we showed that PMap provides more information for orbits close to the central body. That could come from the short integration time of 60 days considered in this work. The objective of this paper is no more an attempt to help to the preparation of a hypothetic future plan for a mission around Apophis, than to carry out a realistic analysis of the interesting problem of celestial mechanics dealing with the dynamical behaviour of a spacecraft around an asteroid undergoing the gravitational effects caused by a close encounter with the Earth.
    \section{Acknowledgements}
       The authors would like to thank the Coordination for the Improvement of Higher Education Personnel (CAPES), which supported this work via the grant 88887.374148/2019-00. VC acknowledge the support of the Brazilian National Research Council (CNPq, grant 301577/2017-0

    \section*{Conflict of interest}
      The authors declare that they have no conflict of interest.

   \bibliographystyle{spbasic}    
   \bibliography{mybib.bib}       

        
   \onecolumn
   \appendix
   \section{Appendix 1: The variations of Apophis axis}\label{appendix_1}
      In this appendix, we present the variations of the precession in longitude ($\Delta\psi$), obliquity ($\Delta\varepsilon$), and rotation angle ($\Delta \omega$) of the asteroid, as presented in \citet{souchay_2018}
   
      \begin{eqnarray*}
         \Delta \psi &=& \frac{3GM_{\bigoplus}}{2a^{3}\omega}H_{d}\int\cos I \bigg(\frac{a}{r}\bigg)^{3}\bigg(1-\cos 2(\lambda-h)\bigg)dt -\\
                     & & \frac{3GM_{\bigoplus}}{2a^{3}\omega}H_{t}\int\bigg[\bigg(\frac{a}{r}\bigg)^{3}\big(2\cos I \cos2(l+g)\big)- \big(1+\cos I)\cos2(\lambda-h-l-g\big) + \\
                     & & \hspace{2cm} \big(1-\cos I)\cos2(\lambda-h+l+g\big)\bigg]dt \\
         \Delta \varepsilon &=& \frac{3GM_{\bigoplus}}{2a^{3}\omega}H_{d}\int\sin I \bigg(\frac{a}{r}\bigg)^{3}\sin2(\lambda-h)dt + K^\prime\int\cos I \bigg(\frac{a}{r}\bigg)^{3}\sin I \sin2(l+g)dt+\\
         & & \frac{3GM_{\bigoplus}}{2a^{3}\omega}H_{t}\int\frac{1}{\sin I}\bigg(\frac{a}{r}\bigg)^{3}\bigg[(1+\cos I)^{2}\sin2(\lambda-h-l-g) + (1-cos I)^{2}\sin2(\lambda-h+l+g)\bigg]dt - \\
         & & \frac{3GM_{\bigoplus}}{2a^{3}\omega}H_{t}\int\frac{\cos I}{\sin I}\bigg(\frac{a}{r}\bigg)^{3}\bigg[(1+\cos I)^{2}\sin2(\lambda-h-l-g) - (1-cos I)^{2} \sin2(\lambda-h+l+g)\bigg]dt\\
         \Delta \omega &=& \frac{2\pi k_{2}}{0.334 * T_{0}}\int\frac{M_{\bigoplus}}{M_{a}}\bigg(\frac{a}{r}\bigg)^{3}\bigg(\sin^{2}\delta-\frac{1}{3}\bigg)
      \end{eqnarray*}
      
      \noindent where $M_{\bigoplus} = 5.972\times10^{24}, M_{a} = 5.310\times10^{10}$ kg are, respectively, the mass of the Earth and the mass of Apophis, $k_{2}$ is the Love number of the asteroid. An arbitrary value of 0.25 was chosen following the theoretical discussions of the static and dynamic Love numbers of asteroids \citep{jacobson_2011, efroimsky_2015}. $T_{0} = 30.4$ h is the nominal value of the rotation period, estimated from the rotation light-curves \citep{pravec_2014}. $a = 37725$ km is the minimum distance Earth-Asteroid during the Close Encounter. $r$ is the distance Apophis-Earth. $l, g$ and $h$ are the Andoyer rotational angles. $I$ and $\lambda$ are, respectively, the obliquity and longitude angle of the Earth centre with respect to the direction of the Earth at its minimum distance (see Fig. \ref{fig01_obliquity_precession}). $\delta$ is the declination of the Earth with respect to Apophis equatorial plane ($\sin \delta = \sin I \sin \lambda$). $omega$ is the spin rate of the asteroid. $H_{d} = \frac{2C-A-B}{2C}$ and $H_{t} = \frac{B-A}{4C}$ are constants related to the dynamical flattening and tri-axiality of Apophis. $A$, $B$ and $C$ are the moments of inertia along the principal axis of the asteroid \citep{aljbaae_2021}. More details on these equations can be found in \citet{souchay_2018, souchay_2014a, souchay_2014b}

\end{document}